%% file: main.tex
\documentclass[12pt]{article}

\input{config/config}
\input{config/acronyms}

\makeglossaries

\begin{document}

\input{config/commands}

\title{Energy conservation and pressure relaxation in an extended two-temperature model for copper with an electron temperature-dependent interaction potential}
\author{Simon Kümmel\thanks{E-mail: simon.kuemmel@fmq.uni-stuttgart.de} \:and Johannes Roth}
\affil{Institute for Functional Matter and Quantum Technologies, University of Stuttgart, Pfaffenwaldring 57, D-70569 Stuttgart}
\maketitle

\begin{abstract}
\noindent An implementation of an electron temperature-dependent interaction potential for copper in a two-temperature model-molecular dynamics framework is presented. An algorithm for enforcing energy conservation when using such an interaction is provided that is needed due to the changing interaction strength with the degree of excitation. Furthermore, focus is put on how to treat the pressure differences due to an electron temperature gradient following laser irradiation. The influence of various extensions is investigated in large-scale two-temperature model molecular dynamics simulations and compared to existing approaches.
\end{abstract}

\section{Introduction}
Laser pulses with pulse duration of the order of the relaxation processes in the irradiated material lead to a strong electronic excitation of the material \cite{Gamaly_2011_FSLMI}. For the material dynamics, Two-Temperature Model (TTM)-Molecular Dynamics(MD) simulations allow for a very accurate description of the motion of the atoms based on an Interaction Potential (IP) that is designed to reproduce the desired properties of the material in question \cite{Schaefer_2002_TTM,Ivanov_2003_TTM}. The TTM, then couples the atomistic motion to a heat conduction equation that describes the electronic excitation and energy transfer from the electrons to the lattice. \\
In the past, a lot of attention has been given to accurate and reliable simulation parameters such as the optical properties at various wavelengths, densities and temperatures \cite{Leng_1998_DCP,Winter_2017_Cu} or the electron heat capacity, heat conductivity and electron-phonon coupling parameter \cite{Lin_2008_heat_capacity_coupling} obtained from experiments and Density Functional Theory (DFT) calculations. At the same time, comparatively little research has been done on the influence of the interatomic interaction, which, as was shown by numerous DFT calculations, does indeed change with the degree of excitation \cite{Recoules_2006_FTDFT,Norman_2012_MDT,Migdal_2015_FTDFT,Daraszewicz_2023_FTMD_Au,Khakshouri_2008_FTMD_W,Murphy_2015_FTMD_W,Tanaka_2022_TTM}. For some materials such as tungsten \cite{Khakshouri_2008_FTMD_W,Murphy_2015_FTMD_W}, gold \cite{Norman_2012_MDT,Daraszewicz_2023_FTMD_Au} and silicon \cite{Shokeen_2010_FTMD_Si,Kiselev_2016_HLRS,Kiselev_2017_Dissertation,Bauerhenne_2024_TTM_MD}, interaction potentials for the use in MD simulations have been developed. \\
The most notable works on copper are the ones by Tanaka et al. \cite{Tanaka_2018_TTM,Tanaka_2022_TTM,Tanaka_2023_TTMMD}. In an implementation of the TTM \cite{Tanaka_2018_TTM}, the relevance of the electron entropy has already been stated and the need for an energy conservation scheme has been proposed \cite{Tanaka_2023_TTMMD}. In the latter work, a combined TTM-MD scheme is used with an electron temperature-dependent potential for copper developed in \cite{Tanaka_2022_TTM}. As previously stated \cite{Kuemmel_2025_MDT}, this potential is capable of reproducing the effect of bond hardening but is not straightforward in its development and is limited in its accuracy. In the works of Tanaka et al. \cite{Tanaka_2022_TTM, Tanaka_2023_TTMMD}, it is still unclear whether or how the change of the reference energy that was used during the potential development is taken into account. Furthermore, it is unclear whether the electron temperatures that are used for the calculation of the interaction potential and the electronic properties are consistent once energy conservation is established. Finally, the density-dependence of the energy conservation scheme has not been described. \\
In this work, an implementation strategy for the establishment of energy conservation during TTM-MD simulations is presented, taking the change of the bond strength and reference energy into account. Furthermore, the correct treatment of pressure differences in electron temperature gradients is explained and compared with a more commonly-used implementation of the blast force \cite{Falkovsky_1999_electron_pressure} as additional force term. In particular, it will be shown that electron temperature-dependent IPs are capable of reproducing the blast force naturally. These investigations are based on the interaction potential developed in \cite{Kuemmel_2025_MDT}. As was the case with the interaction potential, the aim is to present a straightforward, easy-to-implement and well-performing implementation. \\
This work is structured as follows: First, the energy conservation scheme is presented where the origins of its necessity is discussed, a density-dependent implementation is proposed and edge cases are explained. Next, the blast force is introduced and compared to the electron temperature-dependent potentials in a model system. Finally, large-scale TTM-MD simulations are performed where various extensions are systematically compared and the consequences on the material dynamics are demonstrated.

\section{Energy conservation}
\begin{figure}[H]
  \centering
  \begin{subfigure}{0.49 \textwidth}
    \includegraphics[width=0.95 \textwidth]{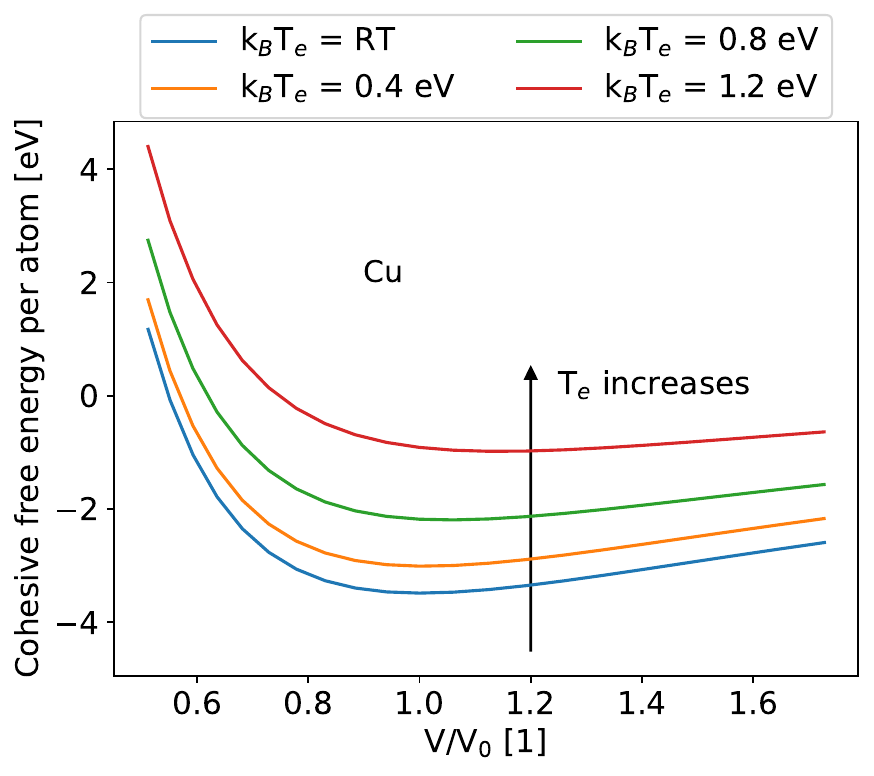}
  \end{subfigure}
  \hfill
  \begin{subfigure}{0.49 \textwidth}
    \includegraphics[width=0.95 \textwidth]{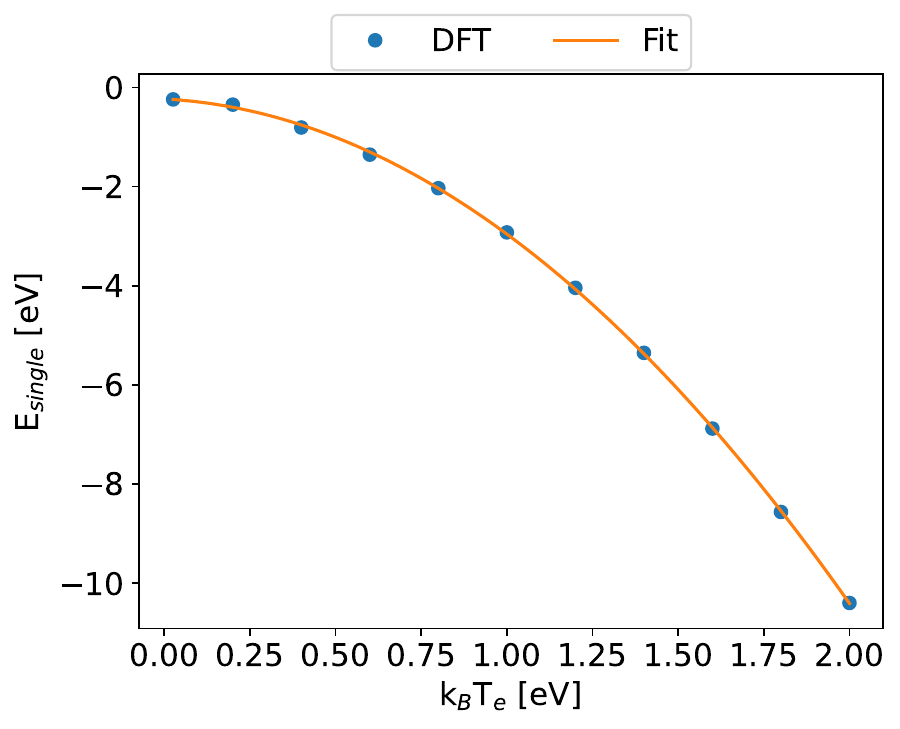}
  \end{subfigure}
  \caption{Left: Cohesive free energy of copper depending on the density at various electron temperatures. Right: Single-atom energy depending on the electron temperature. The single-atom energies are fit to a second-order polynomial with the parameters that are given in table \ref{tab:Fit_parameters}. RT stands for room temperature, i.e. \SI{0.025851}{eV}. Both figures have previously been shown in \cite{Kuemmel_2025_MDT}.}
  \label{fig:Change_of_energy}
\end{figure}

There are two reasons why energy conservation is a relevant aspect when using electron temperature-dependent IPs. Firstly, the interaction strength changes with the degree of excitation, which can be seen in the depth of the potential well or the changing phonon spectra and elastic constants, as shown in \cite{Kuemmel_2025_MDT}. Furthermore, during the potential development, the single-atom energies at the respective electron temperatures were adopted as reference energies. Since these single-atom energies themselves change with the degree of excitation, this has to be taken into account as well. In figure \ref{fig:Change_of_energy}, the electron temperature-dependencies of the lattice free energy and the single-atom energies are visualized.

\begin{figure}[H]
  \centering
  \begin{subfigure}{0.49 \textwidth}
    \includegraphics[width=0.95 \textwidth]{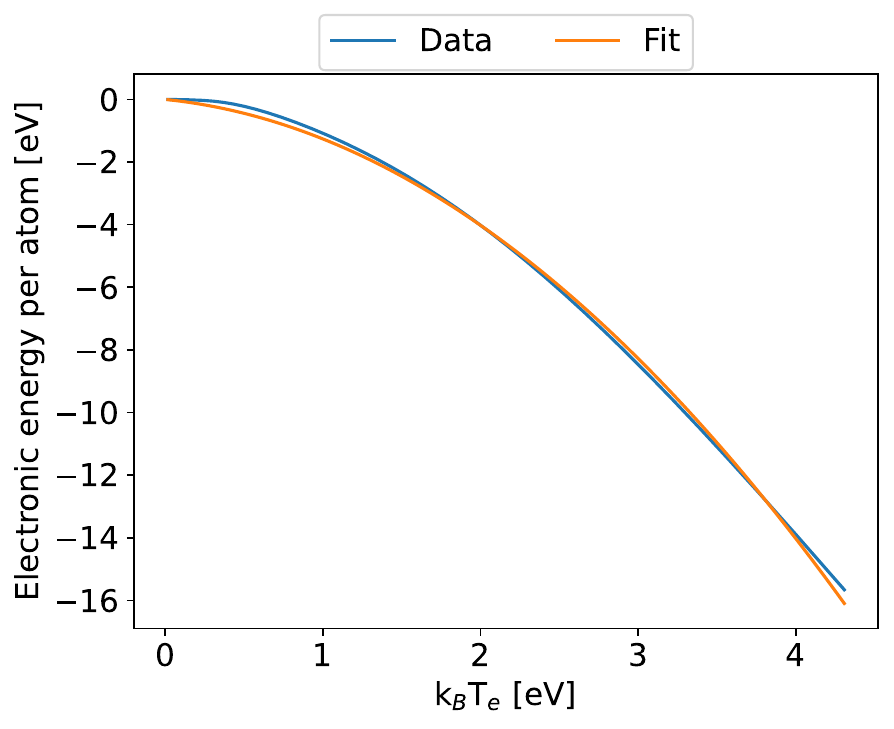}
  \end{subfigure}
  \hfill
  \begin{subfigure}{0.49 \textwidth}
    \includegraphics[width=0.95 \textwidth]{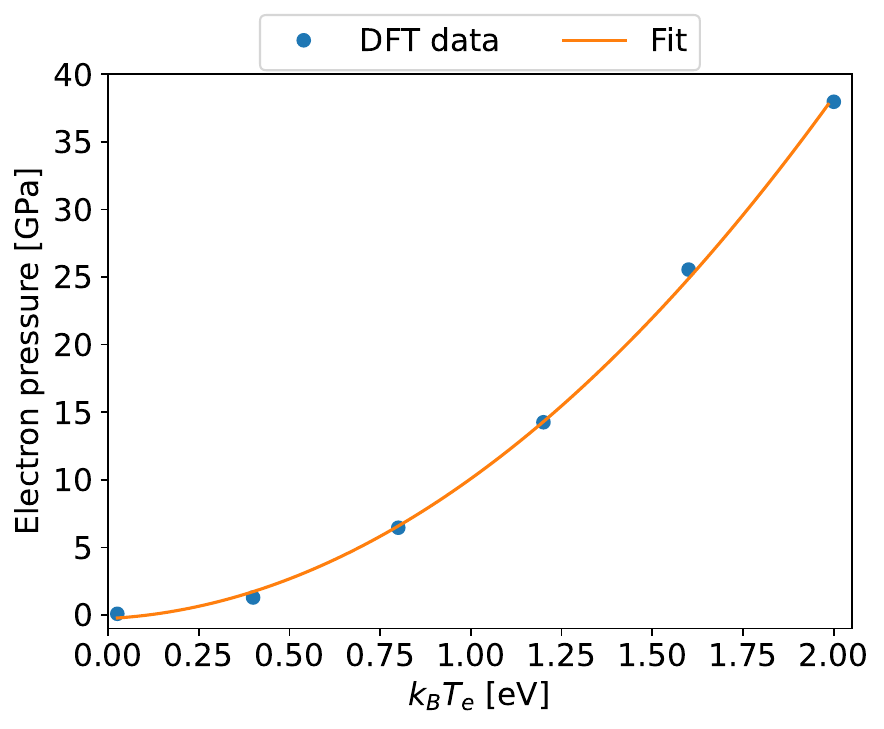}
  \end{subfigure}
  \caption{Left: Electron energy of copper depending on the electron temperature obtained from DFT calculations \cite{Lin_2008_heat_capacity_coupling} and fitted to a second-order polynomial. Right: Electron temperature-dependent electron pressure obtained from DFT calculations \cite{Kuemmel_2025_MDT} with a second-order polynomial fit. The fit parameters of both are given in table \ref{tab:Fit_parameters}.}
  \label{fig:electron_energy_pressure}
\end{figure}

\begin{figure}[H]
  \centering
  \begin{subfigure}{0.32 \textwidth}
    \includegraphics[width=0.99 \textwidth]{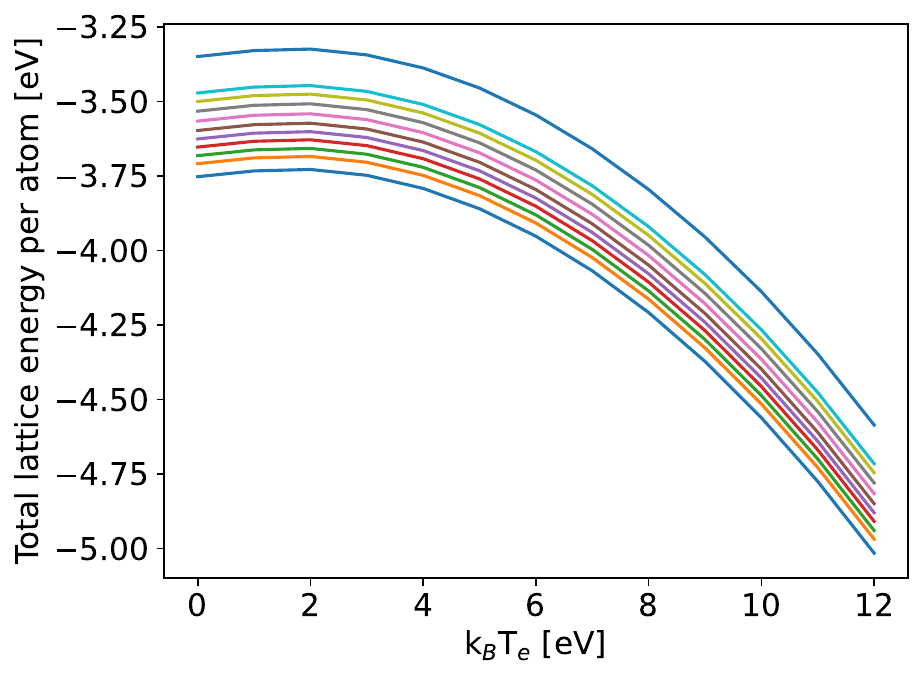}
  \end{subfigure}
  \hfill
  \begin{subfigure}{0.32 \textwidth}
    \includegraphics[width=0.99 \textwidth]{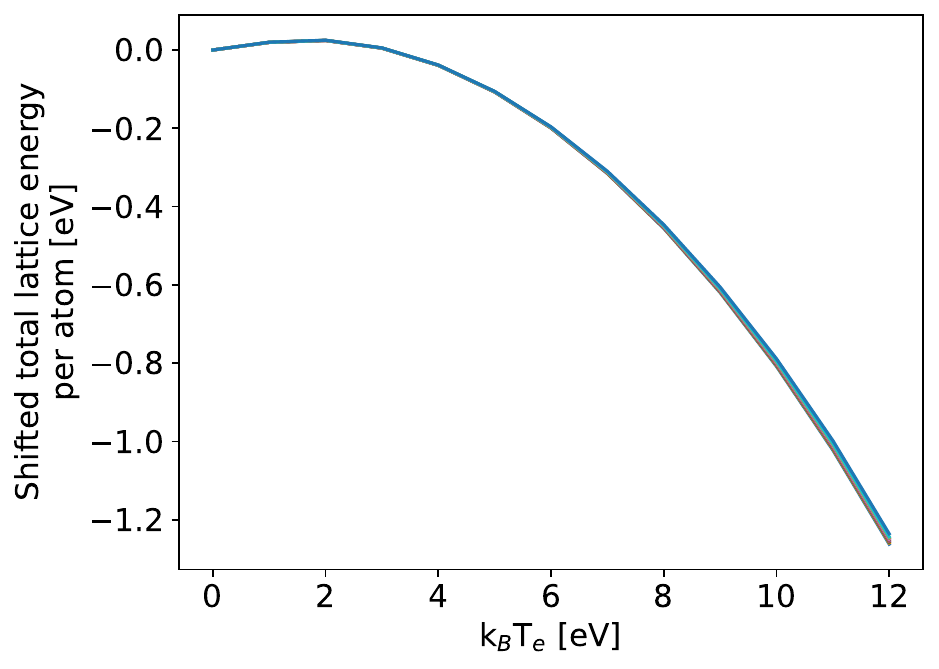}
  \end{subfigure}
  \hfill
  \begin{subfigure}{0.31 \textwidth}
    \includegraphics[width=0.99 \textwidth]{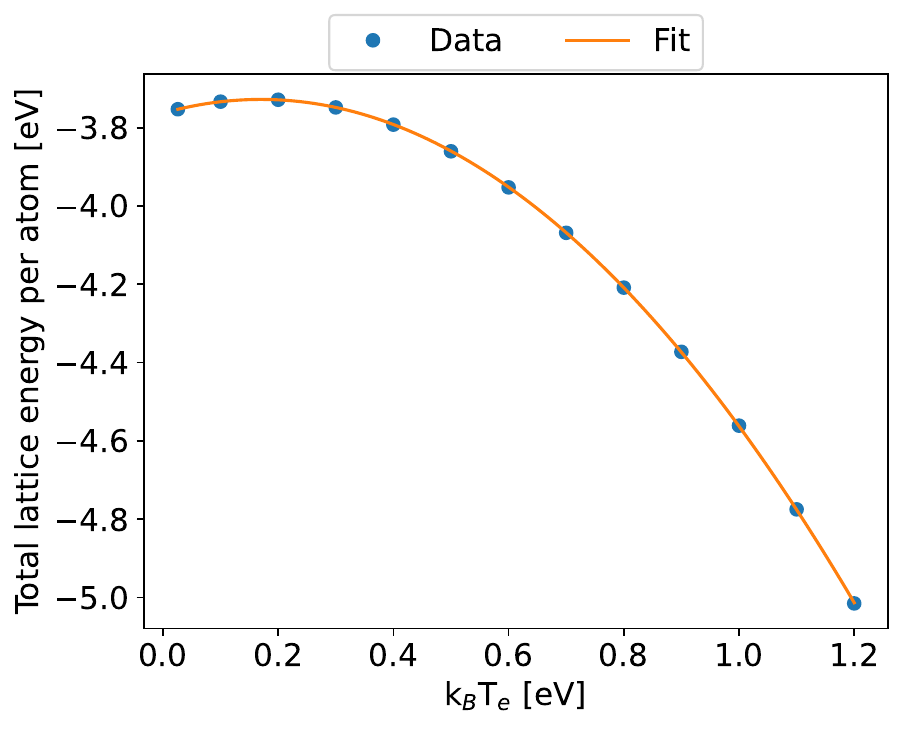}
  \end{subfigure}
  \caption{Left: Total free energy per atom of various structures depending on the electron temperature. Centre: The same as on the left but shifted so that all free energy curves start at \SI{0.0}{eV}. Right: Total free energy per atom in the unexcited state with a fit to a second-order polynomial. In the left and centre figure, each line corresponds to a different structure with varying degree of disorder. The parameters are given in table \ref{tab:Fit_parameters}.}
  \label{fig:total_energy}
\end{figure}

Using the heat capacity data from \cite{Lin_2008_heat_capacity_coupling}, the electron energy $E_{\text{e}}$ can approximately be fitted to a second-order polynomial
\begin{equation}
  E_{\text{e}}\left( T_{e} \right) = \sum_{k=0}^{2} p_{k} \left( \text{k}_{\text{B}} T_{e} \right)^{k} = \lambda \text{k}_{\text{B}} T_{e} + \frac{\gamma}{2} \left( \text{k}_{\text{B}} T_{e} \right)^{2} ,
\end{equation}
which is shown in figure \ref{fig:electron_energy_pressure}. The quadratic dependence of the electron energy on the electron temperature is motivated by the Sommerfeld expansion \cite{Sommerfeld_1928_Sommerfeld}. The linear term in electron temperature was added for a more accurate fit. With the applied fit, the difference between the data and the fit function is only marginal. \\
As was shown previously \cite{Kuemmel_2025_MDT}, the single-atom energy can also be expressed by a second-order polynomial similarly to the electron energy. The parameters of the electron energy fit and the single-atom energy fits are given in table \ref{tab:Fit_parameters}. Realizing that the total free energy
\begin{equation}
  E_{\text{tot}}\left( T_{e} \right) = E_{\text{coh}}\left( T_{e} \right) + E_{\text{single}}\left( T_{e} \right)
\end{equation}
of various lattice configurations show the same overall behaviour apart from a structure-dependent shift $a^{\text{str}}$, one of them was fitted to a second-order polynomial as well. Then, the electron energy reads
\begin{equation}
  E_{\text{tot}} = a_{\text{str}} + b \left( k_{\text{B}} T_{\text{e}} \right) + c \left( k_{\text{B}} T_{\text{e}} \right)^{2} . \label{eq:total_energy}
\end{equation}
The zeroth order parameter $a_{\text{str}}$ doesn't play a significant role, as will become obvious later. In particular, the fit was performed on the energies obtained from a completely disordered structure. This choice was made because during and shortly after laser irradiation, the density is practically unchanged and at ambient density, the overall behaviour of the energy is almost identical with electron temperature apart from the shift $a_{\text{str}}$. For longer times after laser irradiation, the surface starts to melt and, depending on the laser fluence, undergoes ablation. In both scenarios, a disordered structure is more prevalent. The relative difference between the shifted total energy of the solid and liquid structure is about \SI{10}{\percent}. \\
The various free energy curves and the resulting fit are shown in figure \ref{fig:total_energy} at ambient density.

\begin{table}[h]
  \centering
  \caption{Fit parameters of the single-atom free energies and the electron heat capacity. All values are rounded to 8 digits after the decimal point. The units of the parameters given below are $\left[ E_{\text{single}} \right] = \unit{eV}$, $\left[ E_{\text{e}} \right] = \unit{eV/atom}$ and $\left[ p_{\text{e}} \right] = \unit{GPa}$.}
  \label{tab:Fit_parameters}
  \begin{tabular}{c|c|c|c|c}
    Parameter $P_{k}$ & $p_{k, 0}$ $\left[ \frac{[P_{k}]}{\left( \text{k}_{\text{B}} T_{\text{e}} \right)^{0}} \right]$ & $p_{k, 1}$ $\left[ \frac{[P_{k}]}{\left( \text{k}_{\text{B}} T_{\text{e}} \right)^{1}} \right]$ & $p_{k, 2}$ $\left[ \frac{[P_{k}]}{\left( \text{k}_{\text{B}} T_{\text{e}} \right)^{2}} \right]$ \\ \hline
    $E_{\text{single}}$ & -0.23359741       & -0.36458457 & -2.36130430 \\
    $E_{\text{e}}$      & 0.0               & −0.51354432 & -0.74824956 \\
    $p_{\text{e}}$      & -0.25688985       & 1.39621806  & 8.94976771
  \end{tabular}
\end{table}

\begin{figure}[H]
  \centering
  \begin{subfigure}{0.32 \textwidth}
    \includegraphics[width=0.99 \textwidth]{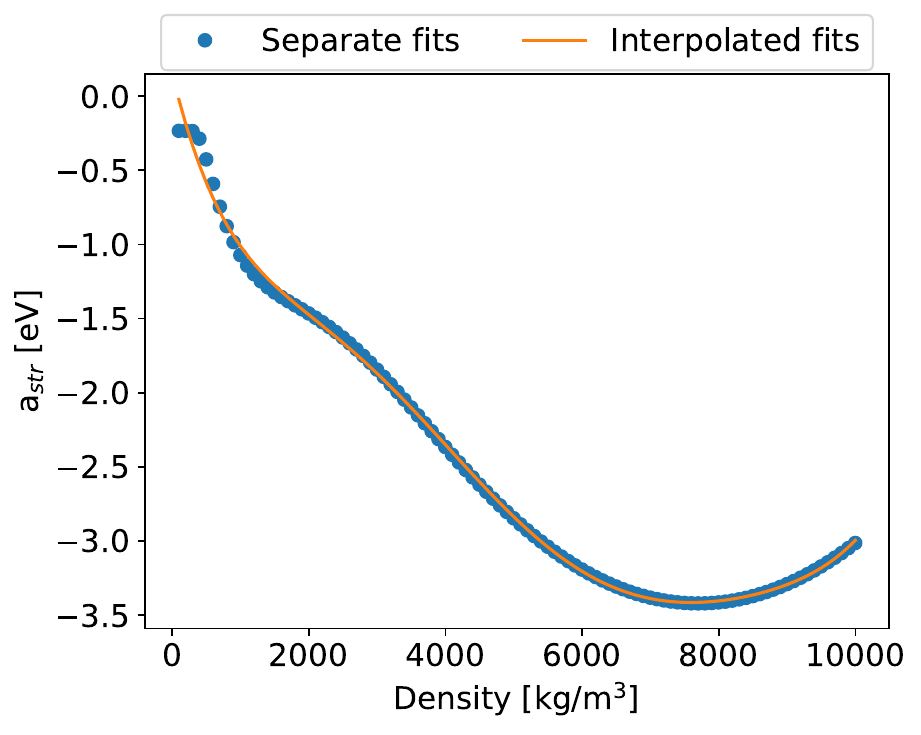}
  \end{subfigure}
  \hfill
  \begin{subfigure}{0.32 \textwidth}
    \includegraphics[width=0.99 \textwidth]{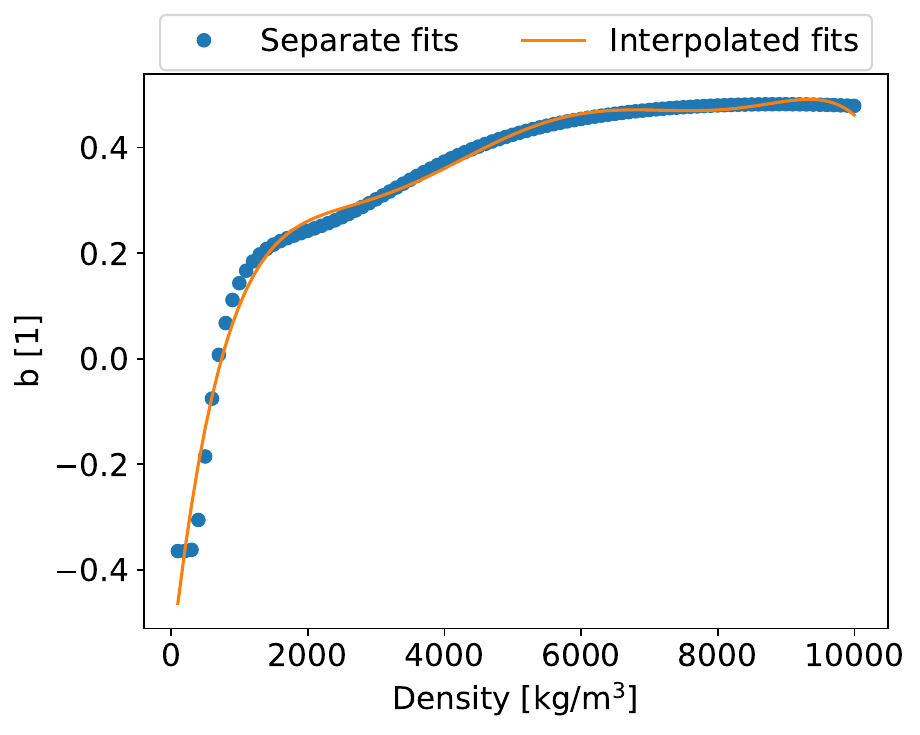}
  \end{subfigure}
  \hfill
  \begin{subfigure}{0.32 \textwidth}
    \includegraphics[width=0.99 \textwidth]{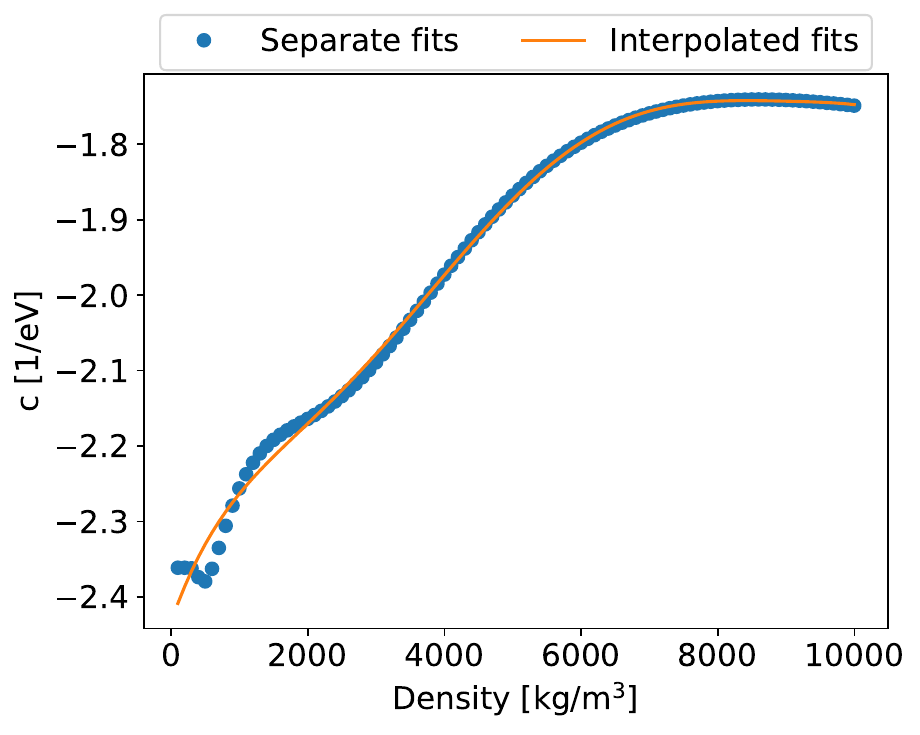}
  \end{subfigure}
  \caption{Density-dependent fit parameters of the lattice energy, which itself depends on the electron temperature. On the left, the fit parameter $p_{k, 0}^{E_{\text{i}}}$, in the centre $p_{k, 1}^{E_{\text{i}}}$ and on the right $p_{k, 2}^{E_{\text{i}}}$ is shown. Each parameter is fitted to a sixth-order polynomial. The parameters of these polynomials are given in table \ref{tab:fit_parameters_2}.}
  \label{fig:fit_parameters}
\end{figure}

Since the lattice energy depends on the density of the material, this procedure was repeated in a large density range between \SI{100}{kg/m^{3}} and \SI{10000}{kg/m^{3}}. The parameters $a_{\text{str}}$, $b$ and $c$ from equation \ref{eq:total_energy} show a clear trend with density, which means that each parameter can again be fitted to a polynomial. For an optimal accuracy, a sixth-order polynomial fit in the density $\rho$ is employed, i.e.
\begin{equation}
  c \left( \rho \right) = \sum_{l=0}^{6} c_{l} \rho^{l}.
\end{equation}
The resulting fit parameters are shown in figure \ref{fig:fit_parameters} and given in table \ref{tab:fit_parameters_2}. There are small deviations between the data and the fit at very low densities. While these deviations will introduce an additional error, it can be assumed to be neglectable since the atoms in regions with such low densities aren't strongly bound anyways and the exact interaction doesn't play a significant role.

\begin{table}[h]
  \centering
  \caption{Fit parameters of the single-atom free energies and the electron heat capacity. All values are rounded to 5 digits after the decimal point. The units of the parameters given below are $\left[ a_{\text{str}} \right] = \unit{eV}$, $\left[ b \right] = \unit{1}$ and $\left[ c \right] = \unit{1/eV}$.}
  \label{tab:fit_parameters_2}
  \begin{tabular}{c|c|c|c}
    Parameter $q_{l}$ & $a_{\text{str}}$ & $b$ & $c$ \\ \hline
    $q_{l, 0}$ $\left[ \frac{[q_{l}]}{\left( \unit{kg/m^{3}} \right)^{0}} \right]$ & \num{1.60501e-1} & \num{-5.74400e-1} & \num{-2.43326e0} \\
    $q_{l, 1}$ $\left[ \frac{[q_{l}]}{\left( \unit{kg/m^{3}} \right)^{1}} \right]$ & \num{-1.92599e-3} & \num{1.17092e-3} & \num{2.58630e-4} \\
    $q_{l, 2}$ $\left[ \frac{[q_{l}]}{\left( \unit{kg/m^{3}} \right)^{2}} \right]$ & \num{1.03781e-6} & \num{-6.60111e-7} & \num{-1.24538e-7} \\
    $q_{l, 3}$ $\left[ \frac{[q_{l}]}{\left( \unit{kg/m^{3}} \right)^{3}} \right]$ & \num{-3.30349e-10} & \num{1.91081e-10} & \num{4.21481e-11} \\
    $q_{l, 4}$ $\left[ \frac{[q_{l}]}{\left( \unit{kg/m^{3}} \right)^{4}} \right]$ & \num{5.15687e-14} & \num{-2.87087e-14} & \num{-6.79160e-15} \\
    $q_{l, 5}$ $\left[ \frac{[q_{l}]}{\left( \unit{kg/m^{3}} \right)^{5}} \right]$ & \num{-3.82904e-18} & \num{2.13759e-18} & \num{5.04907e-19} \\
    $q_{l, 6}$ $\left[ \frac{[q_{l}]}{\left( \unit{kg/m^{3}} \right)^{6}} \right]$ & \num{1.09890e-22} & \num{-6.24160e-23} & \num{-1.41695e-23} \\
  \end{tabular}
\end{table}

\begin{figure}[H]
  \centering
  \begin{subfigure}{0.49 \textwidth}
    \includegraphics[width=0.95 \textwidth]{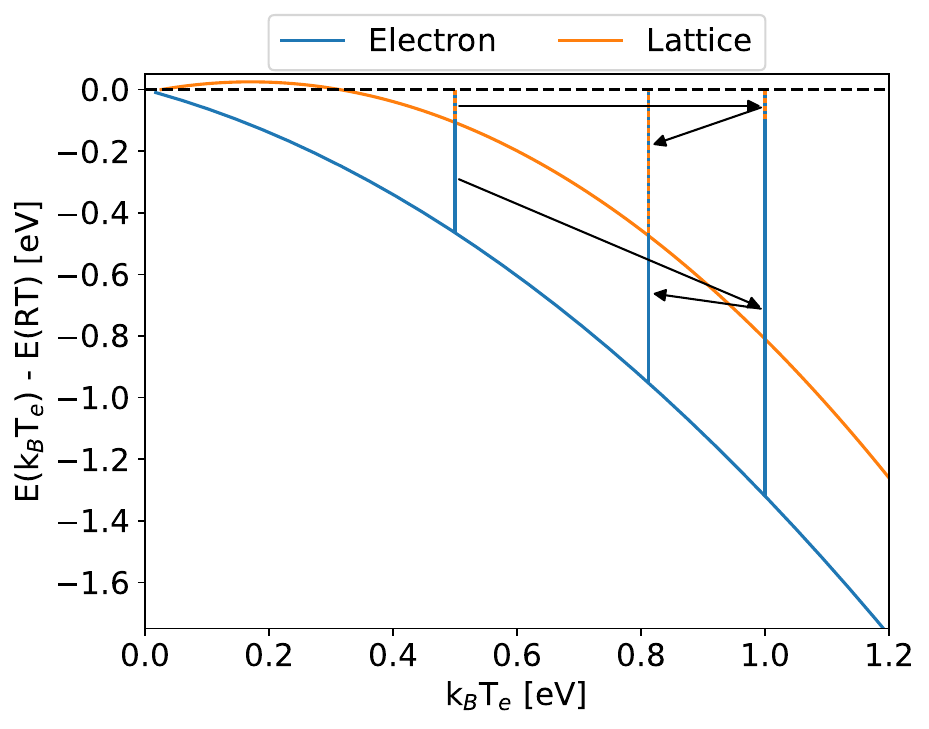}
  \end{subfigure}
  \hfill
  \begin{subfigure}{0.46 \textwidth}
    \includegraphics[width=0.95 \textwidth]{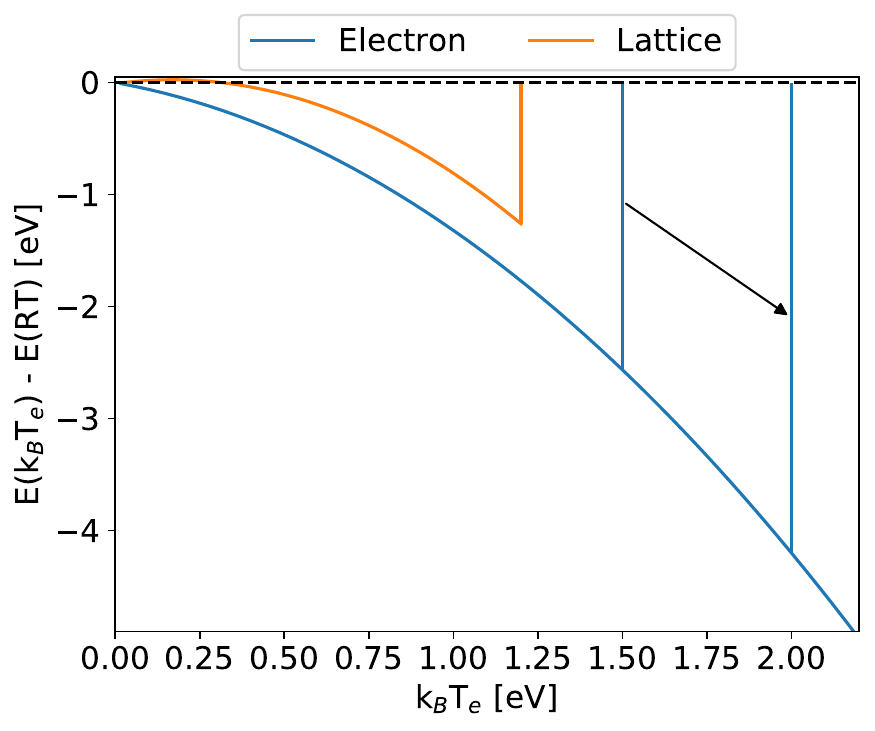}
  \end{subfigure}
  \caption{Visualization of the energy distribution in two situations. On the left, the energy of an FD cell changes due to laser light absorption, heat diffusion or advection from $T_{\text{e}}^{0} = \SI{0.5}{eV}$ to $T_{\text{e}}^{1} = \SI{1.0}{eV}$. After the total energy is distributed in such a way that electron energy and total free energy of the lattice correspond to the same electron temperature, a final electron temperature is found. On the right, the electron of the FD cell before the change of energy is above \SI{1.2}{eV} and the FD cell gains energy. Here, the IP is kept constant and the change of energy is completely compensated by the electron energy.}
  \label{fig:energy_distribution}
\end{figure}

Between two  Finite Differences (FD) timesteps, an arbitrary FD cell is assumed to change its energy due to laser light absorption, heat diffusion or advection. In a first step, it is assumed that the entire energy change $\Delta E$ is compensated by the electron energy, changing the local electron temperature from $T_{\text{e}}^{0}$ to $T_{\text{e}}^{1}$. At this point, the electron temperature that the electron energy and the free energy of the lattice correspond to are different values, which has to be corrected in a second step. For this, $T_{\text{i}}^{0}$ is the electron temperature that is used for the calculation of the initial lattice free energy and $T_{\text{i}}^{\text{f}} = T_{\text{e}}^{\text{f}}$ the lattice and electron temperature that is used for the calculation of the final lattice free energy and electron energy once the energy change is distributed to the electron and lattice free energy. \\
Being able to express all relevant energy contributions to the total energy of the system using second-order polynomials, the energy conservation can be written down as
\begin{align}
  & E_{\text{e}} \left( \text{k}_{\text{B}} T_{\text{e}}^{\text{f}} \right) + E_{\text{i}} \left( \text{k}_{\text{B}} T_{\text{i}}^{\text{f}} \right) - \left[ E_{\text{e}} \left( \text{k}_{\text{B}} T_{\text{e}}^{1} \right) + E_{\text{i}} \left( \min \left( \SI{1.2}{eV}, \text{k}_{\text{B}} T_{\text{i}}^{0} \right) \right) \right] \nonumber \\
  \overset{\hphantom{T_{\text{e}}^{\text{f}} = T_{\text{i}}^{\text{f}} \eqcolon T^{\text{f}}}}{=}& \frac{\gamma}{2} \left( \text{k}_{\text{B}} T_{\text{e}}^{\text{f}} \right)^{2} + \lambda \text{k}_{\text{B}} T_{\text{e}}^{\text{f}} + a^{\text{str}} + b \text{k}_{\text{B}} T_{\text{i}}^{\text{f}} + c \left( \text{k}_{\text{B}} T_{\text{i}}^{\text{f}} \right)^{2} - \frac{\gamma}{2} \left( \text{k}_{\text{B}} T_{\text{e}}^{1} \right)^{2} \nonumber \\
  &- \lambda \text{k}_{\text{B}} T_{\text{e}}^{1} - a^{\text{str}} - b \cdot \min \left( \SI{1.2}{eV}, \text{k}_{\text{B}} T_{\text{i}}^{0} \right) - c \cdot \min \left( \SI{1.2}{eV}, \text{k}_{\text{B}} T_{\text{i}}^{0} \right)^{2} \nonumber \\
  \overset{T_{\text{e}}^{\text{f}} = T_{\text{i}}^{\text{f}} \eqcolon T^{\text{f}}}{=}& \left( \text{k}_{\text{B}} T^{\text{f}} \right)^{2} \underbrace{\left( \frac{\gamma}{2} + c \right)}_{\eqcolon A} + \text{k}_{\text{B}} T^{\text{f}} \underbrace{\left( \lambda + b \right)}_{\eqcolon B} \nonumber \\
  &+ \underbrace{\left[ - \frac{\gamma}{2} \left( \text{k}_{\text{B}} T_{\text{e}}^{1} \right)^{2} - \lambda \text{k}_{\text{B}} T_{\text{e}}^{1} + b \cdot \min \left( \SI{1.2}{eV}, \text{k}_{\text{B}} T_{\text{i}}^{0} \right) - c \cdot \min \left( \SI{1.2}{eV}, \text{k}_{\text{B}} T_{\text{i}}^{0} \right)^{2} \right]}_{\eqcolon \tilde{C}} \nonumber \\
  \overset{\hphantom{T_{\text{e}}^{\text{f}} = T_{\text{i}}^{\text{f}} \eqcolon T^{\text{f}}}}{\overset{!}{=}}& \Delta E
  \label{eq:Energy_balance_1}
\end{align}
\begin{equation}
  \Leftrightarrow A \left( \text{k}_{\text{B}} T_{\text{e}}^{\text{f}} \right)^{2} + B \text{k}_{\text{B}} T^{\text{f}} + \underbrace{\tilde{C} - \Delta E}_{\eqcolon C} = A \left( \text{k}_{\text{B}} T^{\text{f}} \right)^{2} + B \text{k}_{\text{B}} T^{\text{f}} + C \overset{!}{=} 0
\end{equation}
which can then easily be solved using the quadratic formula
\begin{equation}
  \text{k}_{\text{B}} T_{1,2}^{\text{f}} = \frac{-B \pm \sqrt{B^{2} - 4 A C}}{2 A}.
\end{equation}
Only one of the two solutions is the physically meaningful. The condition
\begin{equation}
  \text{k}_{\text{B}} T_{\text{e}}^{\text{f}} > 0 \hspace{0.5 cm}
\end{equation}
can be taken advantage of to determine the meaningful one. Here, the restriction that the electron temperature-dependent potential can only be applied safely at electron temperatures up to \SI{1.2}{eV} was already utilized. Beyond \SI{1.2}{eV}, similar energy balances as the one presented above can be derived. Such cases are detailed in the appendix (appendix \ref{sec:Edge_cases}). A simple example is the case in which the initial electron temperature of an FD cell is above \SI{1.2}{eV} and it absorbs more energy. Then, the total free energy of the lattice is kept constant and the new electron temperature can simply be calculated by the change of electron energy. In figure \ref{fig:energy_distribution}, these two situations are visualized. By summing up the energy contributions after the energy distribution and comparing it to the absorbed energy, it can be shown that the total energy is conserved this way.

\section{Electron pressure}
\begin{figure}[H]
  \centering
  \begin{subfigure}{0.49 \textwidth}
    \includegraphics[width=0.95 \textwidth]{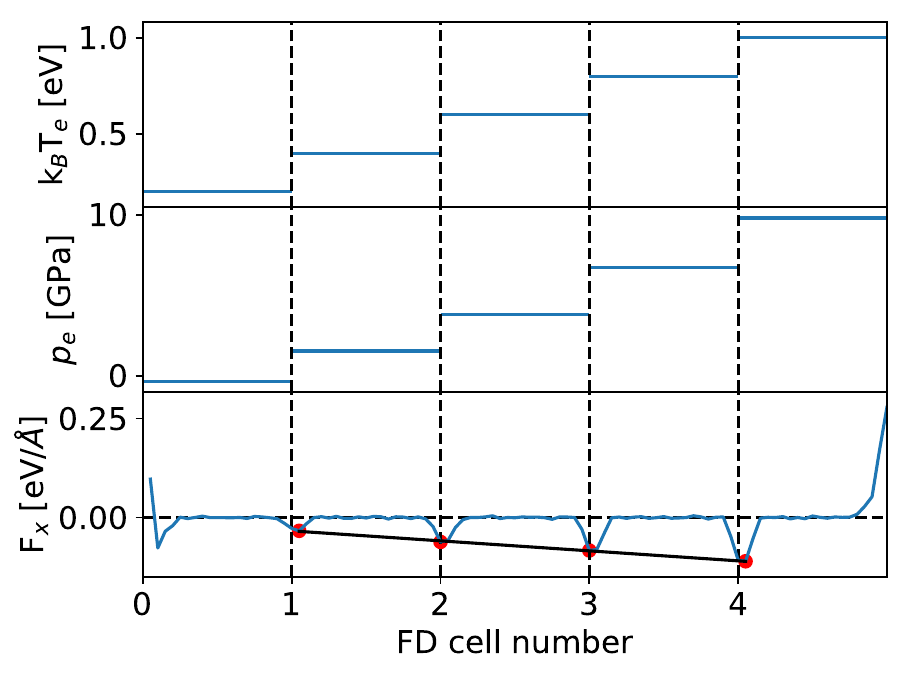}
  \end{subfigure}
  \hfill
  \begin{subfigure}{0.49 \textwidth}
    \includegraphics[width=0.95 \textwidth]{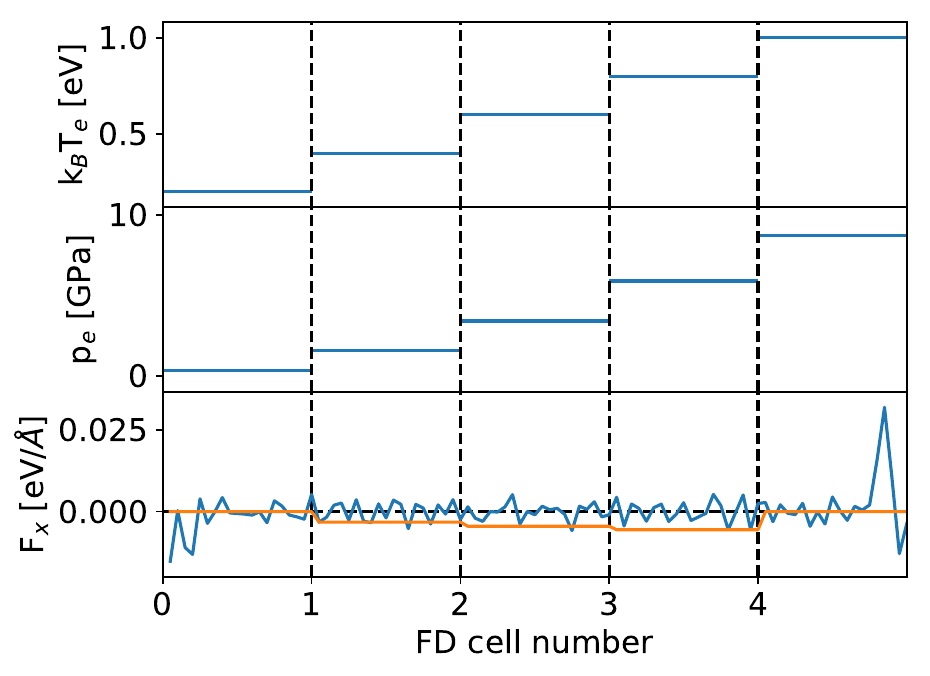}
  \end{subfigure}
  \caption{Top: Electron temperature in five FD cells. Centre: Pressure in each FD cell. Bottom: Force per atom in x-direction. On the left, each of these values is shown for an electron temperature-dependent potential. The red dots indicate the position of maximum force and the black line is a linear fit to these maximum forces according to equation \ref{eq:F_blast_pe}. On the right, the same is shown for an electron temperature-independent potential with additional blast force. Here, an orange curve indicates the additional force contribution due to the electron temperature gradient. In both figures, the spikes in the forces at either end of the sample stem from the open boundary conditions in these simulations}
  \label{fig:temp_press_force}
\end{figure}

Besides the change of the bond strength with electron temperature, the electron temperature-dependent potential also predicts an increase of the electron pressure in a simulation box with fixed volume with increasing electron temperature. Previously, it was shown that the electron temperature-dependent potential correctly reproduces this pressure increase as it was predicted by DFT calculations \cite{Kuemmel_2025_MDT}. In this section, it shall be commented on the effects of pressure gradients due to electron temperature gradients in the sample. \\
As a test system, an oblong simulation box is created that is separated into several FD cells. A constant electron temperature gradient is artificially introduced by explicitly setting the starting electron temperatures in each FD cell. Periodic Boundary Conditions (PBCs) are employed along the axes perpendicular to the electron temperature gradient with open boundary conditions along the axis of the electron temperature gradient. The forces acting on the atoms are collected and averaged within narrow regions along the electron temperature gradient for visualization. The resulting pressures and forces along the electron temperature gradient are shown on the left hand side of figure \ref{fig:temp_press_force}. \\
Without having to develop an electron temperature-dependent IP, several groups \cite{Eisfeld_2018,Winter_2017_Cu,Winter_2017_Cu_2,Yuan_2021_Blast_force,Norman_2013_MDT} have artificially introduced an additional force term, usually referred to as the blast force which is attributed to Falkovsky et al. \cite{Falkovsky_1999_electron_pressure}. This force generally reads
\begin{equation}
  \Vec{F}_{\text{blast}} = - \frac{\Vec{\nabla} p_{\text{e}}}{n_{\text{i}}}.
\end{equation}
where $p_{\text{e}}$ is the electron pressure and $n_{\text{i}}$ is the material density. Assuming
\begin{equation}
  p_{\text{e}} \propto T_{\text{e}}^{2},
\end{equation}
which, based on the data shown in the right part of figure \ref{fig:electron_energy_pressure} seems reasonable, one finds
\begin{equation}
  \Vec{F}_{\text{blast}} \propto 2 T_{\text{e}} \Vec{\nabla} T_{\text{e}}. \label{eq:F_blast_pe}
\end{equation}
As a simple approximation in one dimension, this blast force is implemented as
\begin{equation}
  F_{\text{blast}} \left( x_{i} \right) \approx - \frac{p_{\text{e}} \left(x_{i+1}\right) - p_{\text{e}} \left(x_{i-1}\right)}{2 \left(x_{i+1} - x_{i-1}\right) n_{\text{i}}}. \label{eq:F_blast_IMD}
\end{equation}
In this case, all atoms in FD cell $i$ experience the same force. A comparison of the blast force acting on atom and the forces on the interface using an electron temperature-dependent potential in the previously described electron temperature gradient is shown in figure \ref{fig:temp_press_force}. In both cases, the linear dependence of the blast force on the local electron temperature from equation \ref{eq:F_blast_pe} could be confirmed. Furthermore, simulations with different electron temperature gradients also confirmed the linear dependence of the blast force on this gradient. In the case of the electron temperature-dependent potential in the left part of figure \ref{fig:temp_press_force}, only atoms do indeed experience a force due to the pressure gradient close to the interface of two FD cells with different electron temperatures. However, when explicitly taking the blast force into account in the right part of figure \ref{fig:temp_press_force}, all atoms in an FD cell experience the same force. In both cases, the forces show a peak at both ends of the sample due to the open boundary conditions in these particular simulations. \\

More elaborate implementations of the blast force have been made, such as the one presented in \cite{Norman_2012_MDT,Norman_2013_MDT}. Still, in either implementation, the energy conservation is violated. Furthermore, whenever the blast force is used, assumptions of the density-dependence have to be made which can introduce additional errors. \\

In \cite{Kiselev_2017_Dissertation}, an additional force contribution for atom $i$
\begin{equation}
  F_{i}^{\text{add}} = - \left( \sum_{k} \pdv{V}{P_{k}} \pdv{P_k}{T_{\text{e}}} \right) \dv{T_{\text{e}}}{\Vec{r}_{i}} \approx - \left( \frac{V \left( T_{\text{e}} + \Delta T_{\text{e}} \right) - V \left( T_{\text{e}} \right)}{\Delta T_{\text{e}}} \right) \dv{T_{\text{e}}}{\Vec{r}_{i}}
\end{equation}
was proposed which is supposed to also let the pressure relax for an electron temperature-dependent modified Tersoff potential \cite{Kumagai_2007_Tersoff_MOD}. Here, $V$ is the calculated potential at electron temperature k$_{\text{B}}T_{\text{e}}$ which depends on all potential parameters $P_{k}$.
This expression essentially aims to let the atoms in an FD cell experience the force due to the interaction strength differences in an electron temperature gradient. This ends up giving the same result as the additional blast force term equation \ref{eq:F_blast_pe}, again without considering energy conservation.

\section{TTM-MD simulations}

\subsection{Equations to solve}
In order to demonstrate the influence of the electron temperature-dependent and the importance of the careful implementation of energy conservation, a set of simulations of laser ablation on a copper target are conducted, employing the various models and comparing them. The TTM-MD implementation by \cite{Eisfeld_2022} is used which solves the coupled equations
\begin{equation}
  \pdv{e_{\text{e}} \rho}{t} = \Vec{\nabla} \cdot \left( \kappa \Vec{\nabla} T_{\text{e}} \right) - \Vec{\nabla} \cdot \left( \Vec{u} e_{\text{e}} \rho \right) - G \left( T_{\text{e}} - T_{\text{i}} \right) + Q_{\text{abs}}
\end{equation}
for the electronic subsystem and
\begin{equation}
  m_{i} \dot{\Vec{v_{i}}} = \Vec{F}_{i} + \xi_{i} m_{i} \Vec{v_{i}}
\end{equation}
for the lattice with
\begin{equation}
  \xi_{i} = \frac{\frac{1}{n} \sum_{k}^{n} G V \left( T_{\text{e}}^{k} - T_{\text{i}} \right)}{\sum_{j} m_{j} \left( v_{j}^{\text{therm}} \right)^{2}}.
\end{equation}
In these equations, $n$ is the number of FD steps between two MD steps, $e_{\text{e}} = C_{V} T_{\text{e}} / \rho$ is the electron energy density, $C_{V}$ the electron heat capacity, $\rho$ the ionic density, $\kappa$ the electron heat conductivity, $\Vec{u}$ the drift velocity, $G$ the electron-phonon coupling parameter, $m_{i}$ is the mass, $v_{i}$ the velocity and $\Vec{F}_{i}$ the total force acting on particle $i$. The laser energy is absorbed through an temperature- and density-dependent Lambert-Beer law
\begin{align}
  Q_{\text{abs}} (t, z) &= A I(t) \alpha(z) \exp\left[ - \int_{0}^{z}\alpha (z') \dd{z'} \right] \nonumber \\
  &\approx A I(t) \alpha(z) \exp\left[ - \sum_{i=0}^{n} \alpha(z_{n}) \Delta z \right]
\end{align}
in which the absorption coefficient can change during laser irradiation. In the last expression, the integral is replaced by a sum over $n$ FD cells with a width of $\Delta z$ that the laser light has to travel through to reach the depth $z$. $A = 1 - R$ is the absorbed portion of the total incident intensity $I(t)$ which, here, is modelled as a Gaussian beam
\begin{equation}
  I(t) = I_{0} \exp\left[ -4 \ln(2) \left( \frac{t - t_{0}}{\tau_{\text{p}}} \right)^{2} \right]
\end{equation}
where $\tau_{\text{p}}$ is the Full Width at Half Maximum (FWHM) of the laser and $t_{0}$ is a shift in time so that the laser beam starts at a well-defined time and is calculated via
\begin{equation}
  t_{0} = \tau_{\text{p}} \sqrt{- \frac{\ln\left(\sigma_{\text{th}}\right)}{4 \ln(2)}} + \SI{100}{fs}
\end{equation}
where $\sigma_{\text{th}} = \num{1.0e-5}$ is the threshold intensity at which the laser is considered to be active. The additional \SI{100}{fs} are added as a safety margin. The peak intensity $I_{0}$ is linked to the incident fluence $F_{\text{inc}}$ via
\begin{equation}
  I_{0} = \frac{2 \sqrt{\ln(2)} F_{\text{inc}}}{\sqrt{\pi} \tau_{\text{p}}}.
\end{equation}
$z$ is the depth of a TTM cell from the surface and $d$ is the film thickness. \\
Note that ballistic electron motion \cite{Rethfeld_2017,Chen_2006_TTM} is not taken into account in this work but will be subject to future work. This effect influences the absorption of the laser light and can affect the ablation depth. Therefore, the amounts of ablated material obtained in this work should only be subject to qualitative discussion.

\subsection{Simulation parameters}
The electronic heat capacity and electron-phonon coupling parameter from \cite{Lin_2008_heat_capacity_coupling} are adopted, which both are assumed to be linearly dependent on the density. For the electron heat conductivity, the parameterization by Petrov et al. \cite{Petrov_2016_heat_conductivity} is used that is described in more detail in the supplementary material of that paper. The only changes made to the parameterization are the completely relaxed density at zero temperature $\rho_{0} = \SI{8761}{kg/m^{3}}$, the density at room temperature $\rho_{\text{rt}} = \SI{8628}{kg/m^{3}}$ and the melting temperature at ambient density $T_{\text{m}} = \SI{1168}{K}$ that were calculated with the unexcited potential \cite{Kuemmel_2025_MDT}. These changes have to be made in order to be consistent with the IP that is employed for the following simulations. \\
For the optical properties, the complex dielectric function from the Drude Critical Point model by Winter et al. \cite{Winter_2017_Cu,Winter_2017_Cu_2} which is based on the work by Ren et al. \cite{Ren_2011_epsilon} are used. To be consistent with the collision frequencies used for the calculations of the heat conductivity in \cite{Petrov_2016_heat_conductivity}, the same values are utilized in the calculation of the optical properties. The last change made here is the number of free electrons, which are based on the investigation by Bévillon et al. \cite{Bevillon_2014_free_electrons}. It is assumed that the number of free electrons to be constant up to \SI{0.5}{eV} with a value of 1.9, then increase linearly with the electron temperature with a slope of \SI{0.1485}{eV^{-1}}. \\

Since some of the contributions to the heat conductivity depend on the state of the material, a method of determining whether the material in each FD cell is solid or liquid has to be employed. In order to avoid any inconsistencies between existing phase diagrams, e.g. from \cite{Autrique_2013_phase_diagram} with the here-used IPs -- in particular at high degrees of excitation -- , a  Local Order Parameter (LOP) in each FD cell is calculated. As was presented in \cite{Hoyt_2001_LOP}, for each neighbouring atom, the connecting vector to a central atom is compared to the theoretical connecting vectors of the given structure, here an Face-Centered Cubic (FCC) structure. The lowest differences between the actual and theoretical connecting vectors are added up to give the LOP
\begin{equation}
  \varphi_{i} = \frac{1}{N} \sum_{j=0}^{N} \left| \Vec{r}_{ij} - \Vec{r}^{\text{ FCC, best}} \right|^{2}
\end{equation}
of atom $i$. The approach described in \cite{Hoyt_2001_LOP} was adapted slightly to work with the cell list-based molecular dynamics framework by not considering the $N = 12$ closest atoms but all atoms that are within the radius of
\begin{equation}
  d_{\text{cutoff}} = \left( \frac{\sqrt{2} a_{0}}{2} + a_{0} \right) / 2 \label{eq:d_cutoff}
\end{equation}
which is exactly in the middle between the nearest neighbour distance and the next-nearest neighbour distance. It was ensured that this change has no significant effect on the resulting order parameter. A comparison of the two methods is given in the appendix (appendix \ref{sec:LOP}). The LOP in an FD cell is obtained by averaging over the values of each atom in that FD cell. As in \cite{Hoyt_2001_LOP}, LOPs above $\varphi_{i} = \SI{0.7}{\text{\AA}^{2}}$ indicate that the material is liquid, below that value, the material is solid. \\

For the simulations with explicit blast force, the additional force contribution is implemented as in equation \ref{eq:F_blast_IMD}. The temperature-dependent electron pressure is taken from the data previously obtained \cite{Kuemmel_2025_MDT}. The electron pressure is assumed to depend linearly on the density.

\subsection{Simulation setup}
The sample to be investigated consists of 2000 unit cells in the direction of the laser irradiation and 20 unit cells perpendicular to the direction of laser irradiation. The simulation box extends \SI{100}{nm} beyond the surfaces on either side of the sample. Before irradiating the sample with a laser pulse, it is ensured that no residual pressure is present in the simulation box. This is done by applying a set of equilibration steps containing a sample heatup to room temperature using PBCs in all three directions and an equilibration with open boundary conditions in the direction of the laser. \\
After the sample equilibration, a \SI{1056}{nm} laser pulse with a duration of $\tau_{\text{FWHM}} = \SI{680}{fs}$ and a fluence of $F = \SI{3.8}{J/m^{2}}$ is applied. For the TTM, the sample is divided into 512 FD cells, which are only active if a minimum amount of 100 atoms is in the respective cell. \\

Two sets of simulations are performed, one for a total time of \SI{5}{ps} where the focus is on the energy distribution between the electronic and the atomistic systems and one for a total time of \SI{200}{ps} where the focus is on the ablation time, depth and general behaviour.

\subsection{Simulation results}
\begin{figure}[H]
  \centering
  \includegraphics[width=0.6 \textwidth]{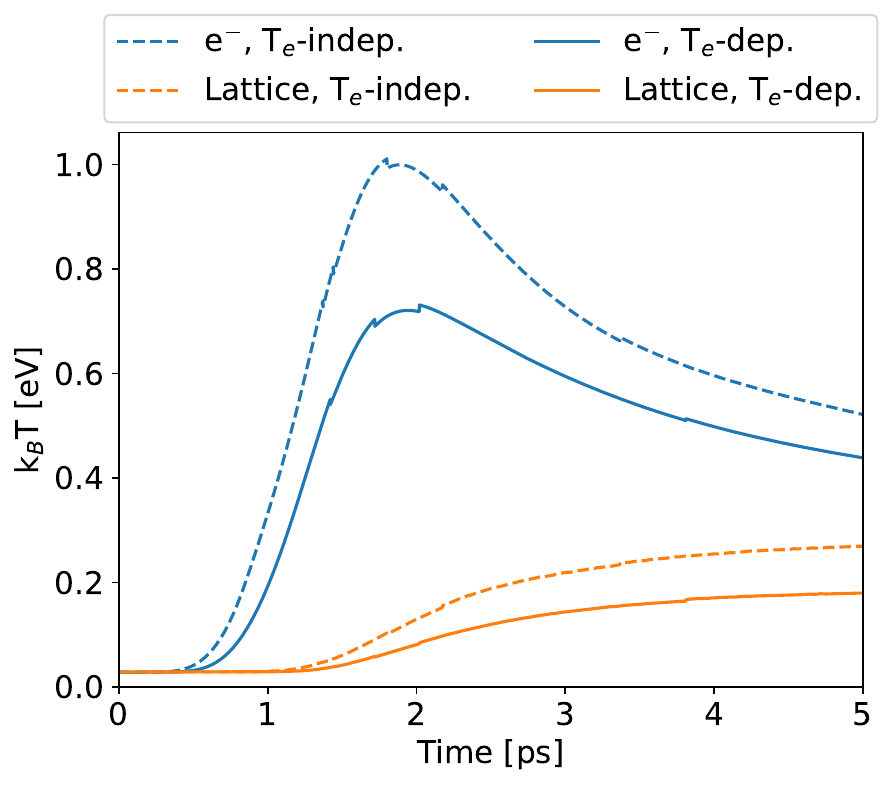}
  \caption{Electron and lattice energy within the skin depth of the material depending on time using an electron temperature-dependent IP during a TTM-MD simulation with and without energy considering energy conservation. The spikes in the temperature stem from the activation or deactivation of FD cells.}
  \label{fig:TTM_temperatures_surface}
\end{figure}

\begin{figure}[H]
  \centering
  \begin{subfigure}{0.49 \textwidth}
    \includegraphics[width=0.95 \textwidth]{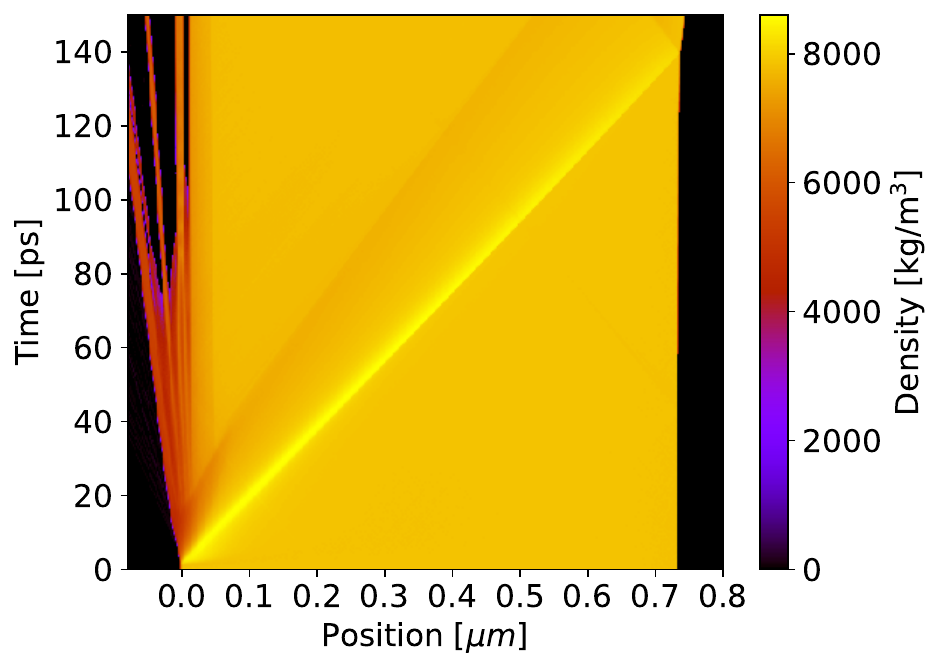}
  \end{subfigure}
  \hfill
  \begin{subfigure}{0.49 \textwidth}
    \includegraphics[width=0.95 \textwidth]{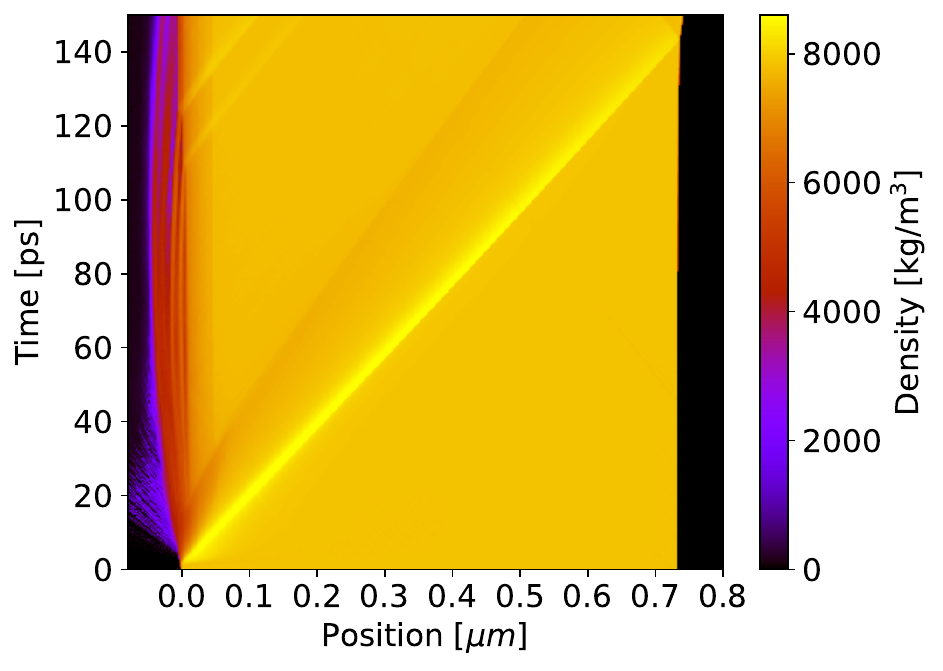}
  \end{subfigure}
  \begin{subfigure}{0.49 \textwidth}
    \includegraphics[width=0.95 \textwidth]{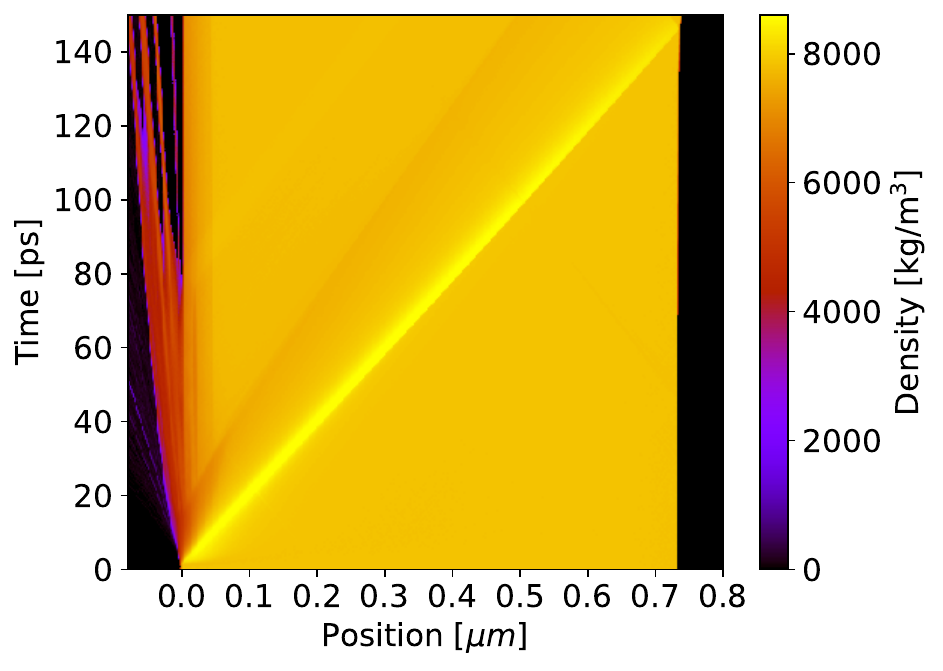}
  \end{subfigure}
  \hfill
  \begin{subfigure}{0.49 \textwidth}
    \includegraphics[width=0.95 \textwidth]{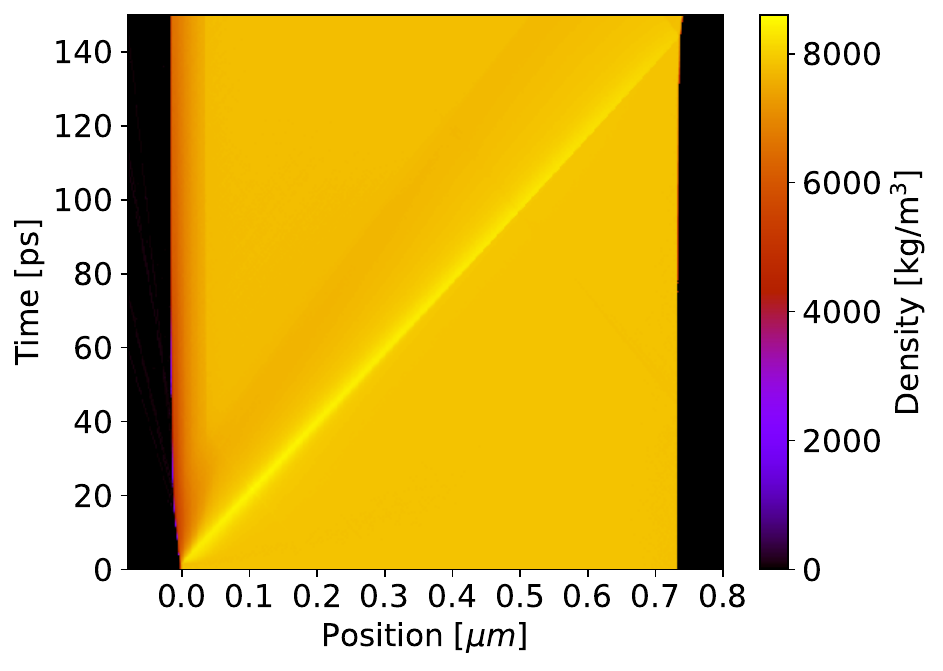}
  \end{subfigure}
  \caption{Temporal and spatial evolution of the density during and after laser irradiation. In the top left figure, the evolution during a simulation using an electron temperature-independent potential is shown. In the top right, the same is shown with an additional force term due to the blast force. In the bottom figures, an electron temperature-dependent potential was utilized. In the bottom left, energy conservation wasn't enforced and in the bottom right, the energy was conserved. Note that all axes are the same for easier comparability.}
  \label{fig:dens_ANA_vs_ANATE_200ps}
\end{figure}

Applying the energy conservation scheme to an actual TTM-MD simulation, it can be observed that the electron temperature is significantly lower than not using it. This can particularly be seen in the temporal evolution of the electron and lattice energy within the skin depth of the material shown in figure \ref{fig:TTM_temperatures_surface}. The lower electron temperature yields a lower electron-phonon coupling parameter which in turn results in a slower increase of the lattice temperature. Also, since the heat transfer from the electrons to the lattice depends on the difference between the respective temperatures, the lattice temperature increases more slowly. This, combined with the fact that the electron temperature-dependent potential predicts a higher melting point at higher electron temperatures, reduces the amount of ablated material and delays the ablation process. This is supported by the density evolutions presented in figure \ref{fig:dens_ANA_vs_ANATE_200ps} which will be discussed in more detail below. \\
Note that the spikes in figure \ref{fig:TTM_temperatures_surface} stem from the activation and deactivation of FD cells. This activation and deactivation is generally needed to ensure the stability of the TTM scheme. Cells with only few atoms that have a very high velocity would otherwise be assigned a very high temperature which can greatly affect light absorption or heat transfer \cite{Eisfeld_2020_Dissertation}.

\begin{figure}[H]
  \centering
  \begin{subfigure}{0.49 \textwidth}
    \includegraphics[width=0.95 \textwidth]{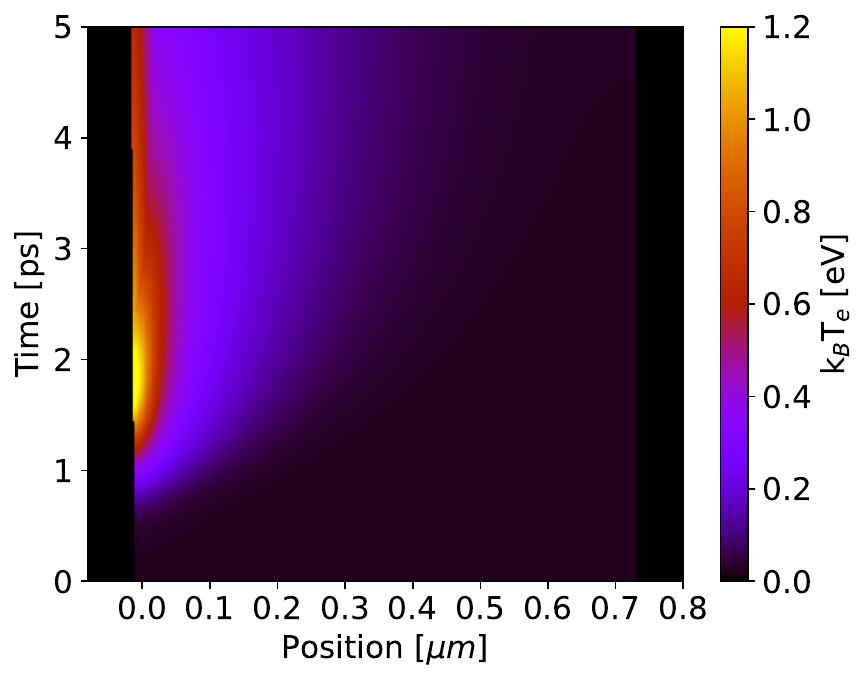}
  \end{subfigure}
  \hfill
  \begin{subfigure}{0.49 \textwidth}
    \includegraphics[width=0.95 \textwidth]{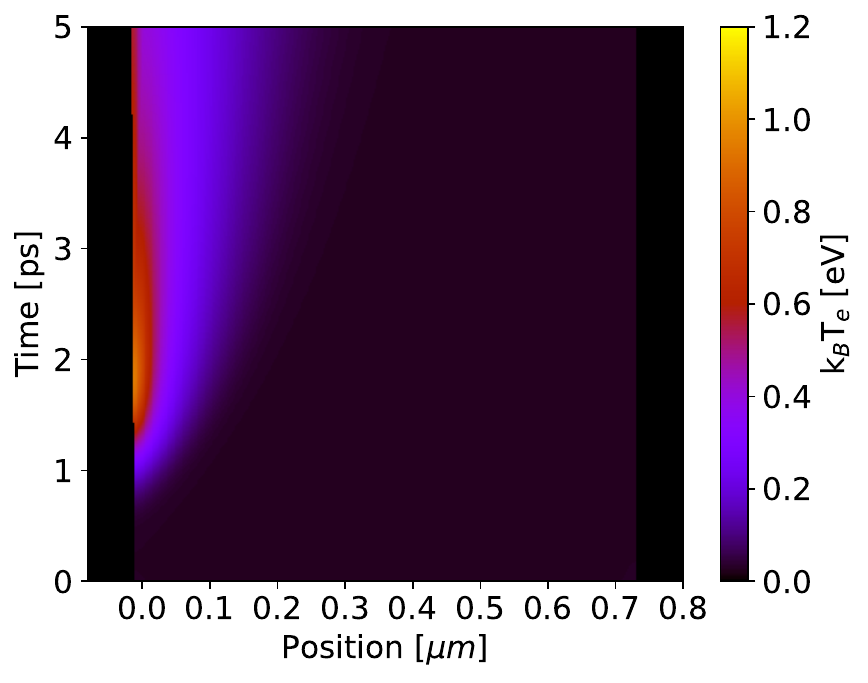}
  \end{subfigure}
  \caption{Temporal and spatial evolution of the electron temperature during and after laser irradiation. On the left, the evolution during a simulation using an electron temperature-independent potential is shown. The same evolution is obtained when employing an additional force term to take blast force into account or when using an electron temperature-dependent potential but not considering energy conservation. On the right, the electron temperature evolution is shown when employing an electron temperature-dependent potential and taking energy conservation into account. Note that all axes are the same for easier comparability.}
  \label{fig:Te_ANA_vs_ANATE_5ps}
\end{figure}

In figure \ref{fig:Te_ANA_vs_ANATE_5ps}, the evolution of the electron temperature in time and space is shown using various IPs. The left figure was obtained using an electron temperature-independent potential with and without the additional blast force and using an electron temperature-dependent potential without taking energy conservation into account. The evolution of the electron temperature is the same in all of these cases since the interaction itself doesn't affect the absorption, heat conduction or transfer of energy from the electronic system to the lattice unless energy conservation is enforced. This particular situation is shown in the right figure of figure \ref{fig:Te_ANA_vs_ANATE_5ps}. \\

\begin{figure}[H]
  \centering
  \begin{subfigure}{0.49 \textwidth}
    \includegraphics[width=0.95 \textwidth]{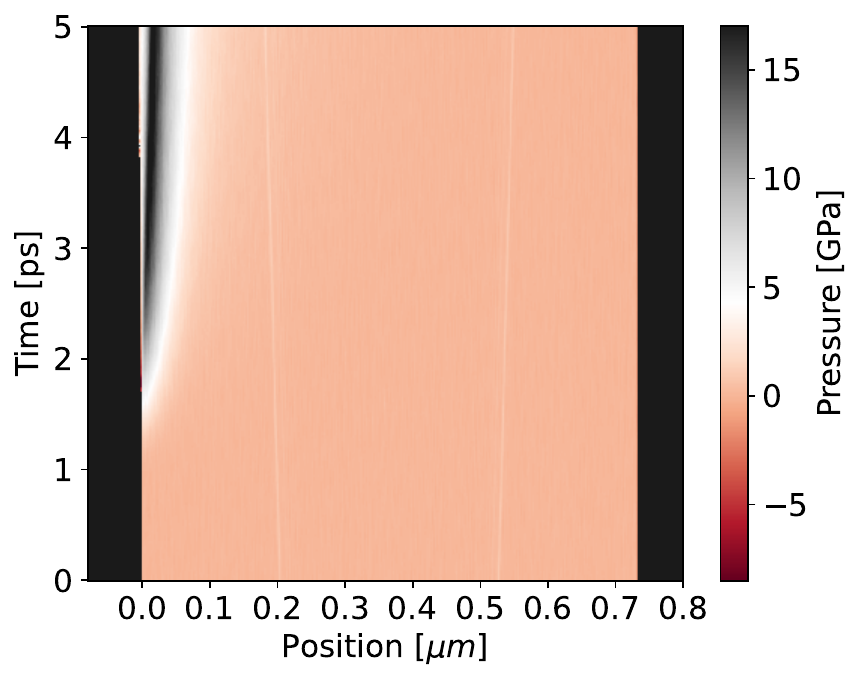}
  \end{subfigure}
  \hfill
  \begin{subfigure}{0.49 \textwidth}
    \includegraphics[width=0.95 \textwidth]{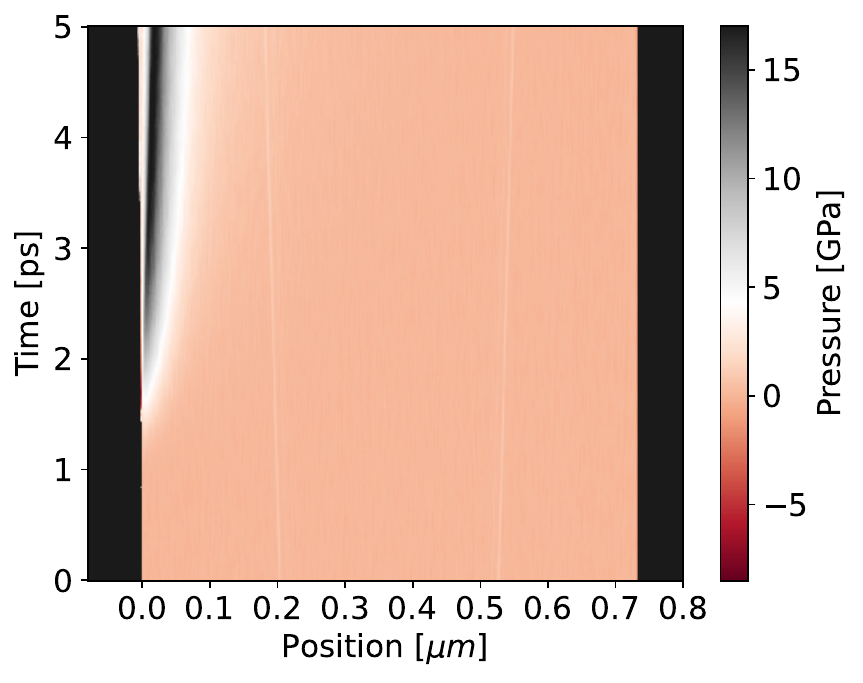}
  \end{subfigure}
  \begin{subfigure}{0.49 \textwidth}
    \includegraphics[width=0.95 \textwidth]{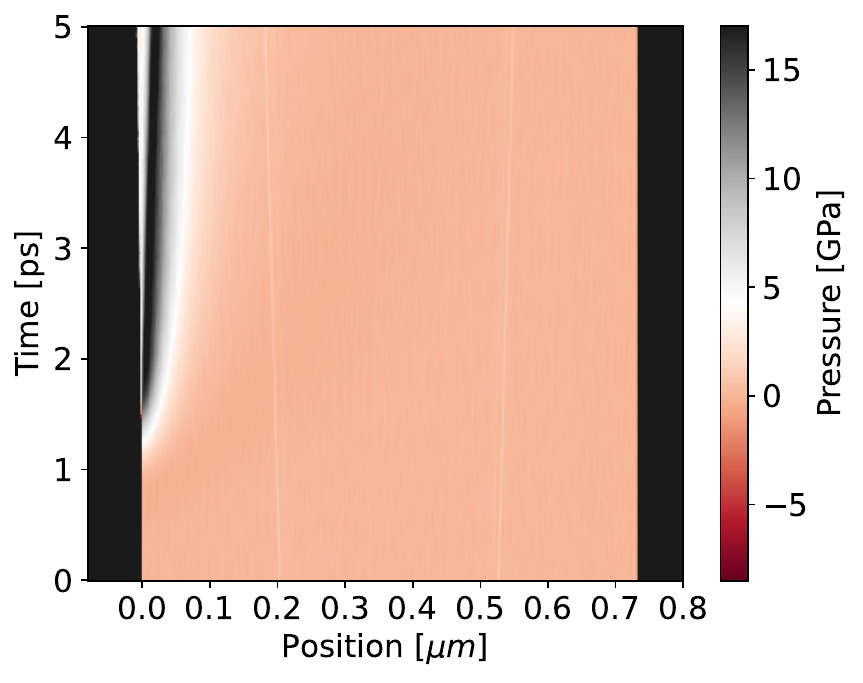}
  \end{subfigure}
  \hfill
  \begin{subfigure}{0.49 \textwidth}
    \includegraphics[width=0.95 \textwidth]{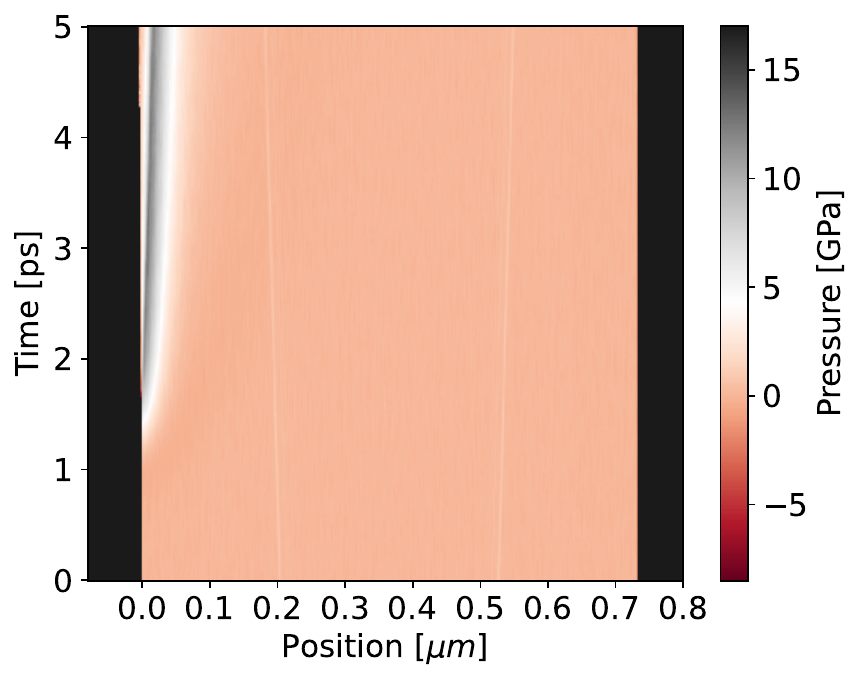}
  \end{subfigure}
  \caption{Temporal and spatial evolution of the pressure during and after laser irradiation. In the top left figure, the evolution during a simulation using an electron temperature-independent potential is shown. In the top right, the same is shown with an additional force term due to the blast force. In the bottom figures, an electron temperature-dependent potential was utilized. In the bottom left, energy conservation wasn't enforced and in the bottom right, the energy was conserved. Note that all axes are the same for easier comparability. Also note that the dark regions at either end of the sample indicate the surfaces.}
  \label{fig:press_ANA_vs_ANATE_5ps}
\end{figure}

Furthermore, the temporal and spatial evolutions of the pressure using various interactions are shown in figure \ref{fig:press_ANA_vs_ANATE_5ps}. In the top left, an electron temperature-independent potential was used. In the top right, the blast force was additionally taken into account. In the bottom left, an electron temperature-dependent potential that doesn't conserve energy was used and in the bottom right, energy was additionally conserved. \\
The additional blast force leads to single atoms being removed from the surface due to the high pressure gradients that occur there. Using the electron temperature-dependent IP on the other hand, the pressure gradients at the surface are less pronounced and no atoms are removed from the surface immediately. Instead, the strength of the pressure wave moving into the sample is increased compared to the electron temperature-independent potential. Finally, the sample experiences a weaker pressure wave moving into the sample using an electron temperature-dependent potential and enforcing energy conservation. This is due to the decreased electron temperature as some of the absorbed energy is used up to balance the changing interaction strength.

\begin{figure}[H]
  \centering
  \begin{subfigure}{0.49 \textwidth}
    \includegraphics[width=0.95 \textwidth]{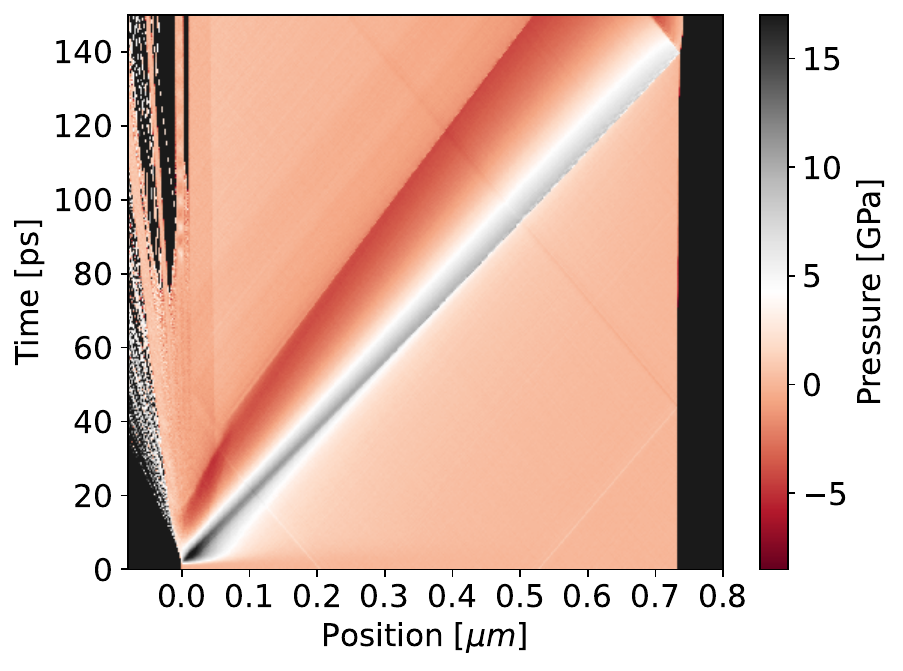}
  \end{subfigure}
  \hfill
  \begin{subfigure}{0.49 \textwidth}
    \includegraphics[width=0.95 \textwidth]{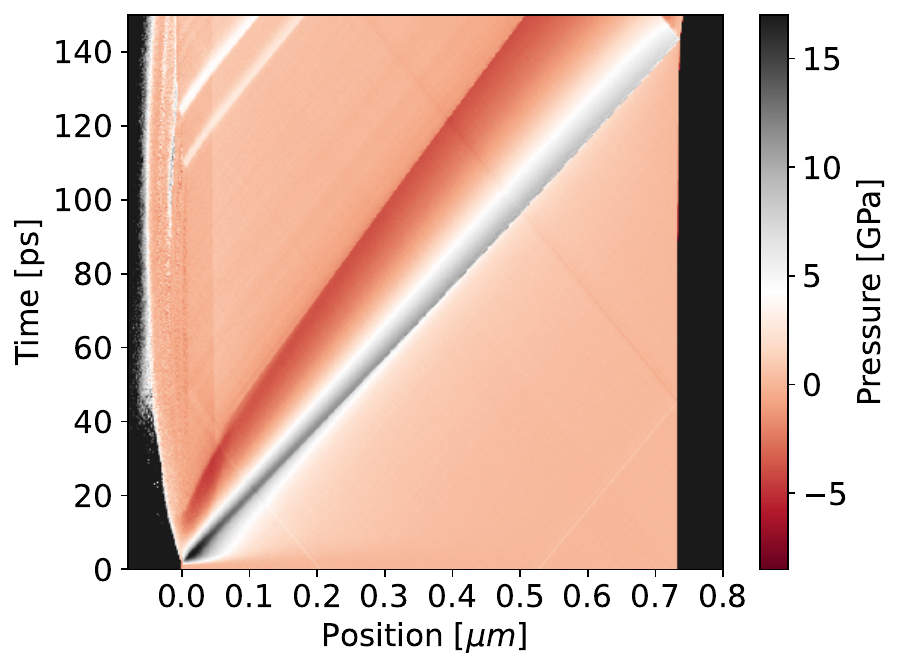}
  \end{subfigure}
  \begin{subfigure}{0.49 \textwidth}
    \includegraphics[width=0.95 \textwidth]{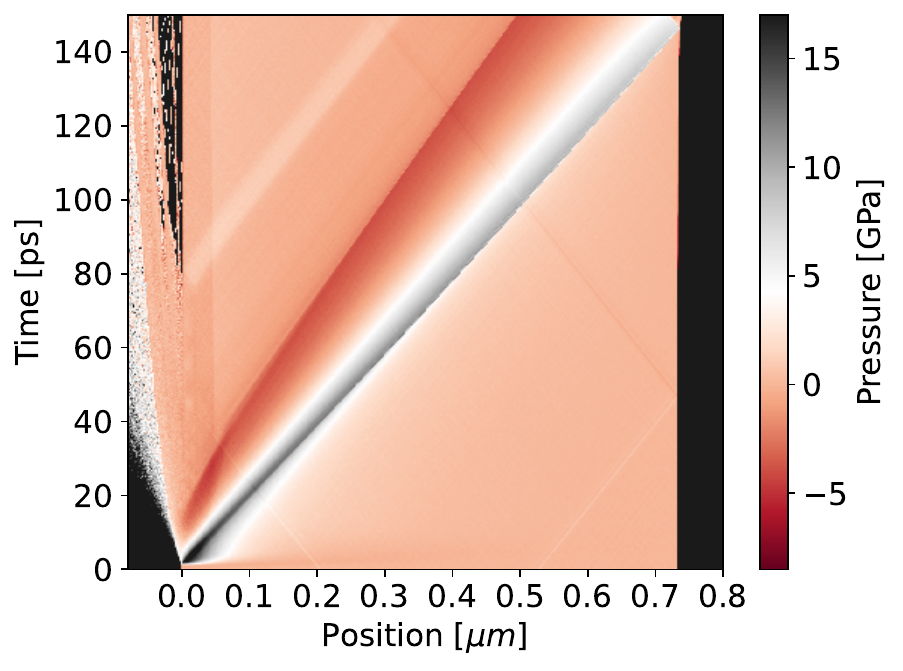}
  \end{subfigure}
  \hfill
  \begin{subfigure}{0.49 \textwidth}
    \includegraphics[width=0.95 \textwidth]{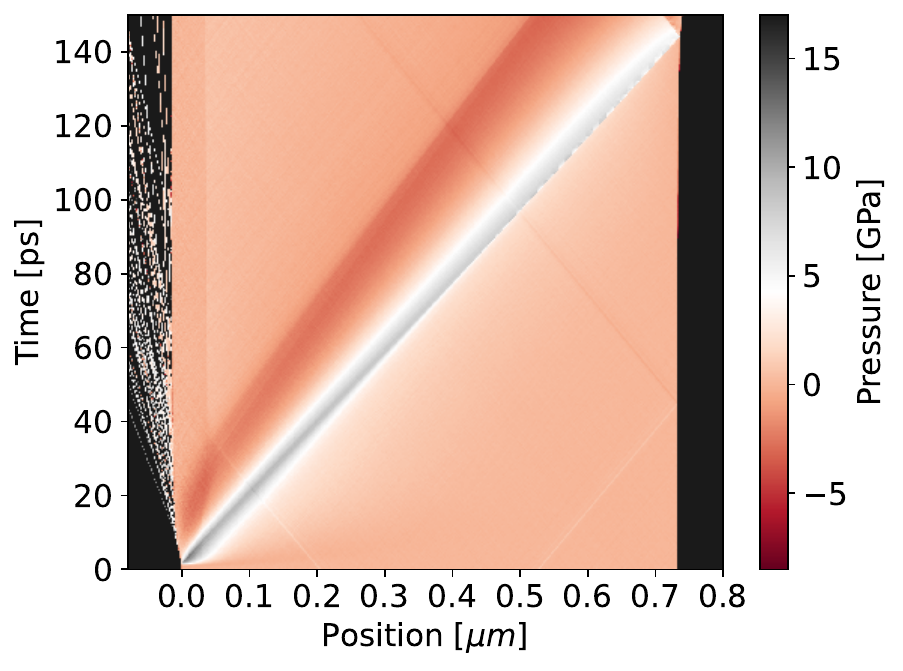}
  \end{subfigure}
  \caption{Temporal and spatial evolution of the pressure during and after laser irradiation. In the top left figure, the evolution during a simulation using an electron temperature-independent potential is shown. In the top right, the same is shown with an additional force term due to the blast force. In the bottom figures, an electron temperature-dependent potential was utilized. In the bottom left, energy conservation wasn't enforced and in the bottom right, the energy was conserved. Note that all axes are the same for easier comparability. Also note that the dark regions at either end of the sample indicate the surfaces.}
  \label{fig:press_ANA_vs_ANATE_200ps}
\end{figure}

In figure \ref{fig:dens_ANA_vs_ANATE_200ps}, the density evolutions of the copper sample obtained with the various interactions are shown. Comparing the simulation with the electron temperature-dependent potential without energy conservation shown at the bottom left of figure \ref{fig:dens_ANA_vs_ANATE_200ps} with the electron temperature-independent potential shown in the top right figure of figure \ref{fig:dens_ANA_vs_ANATE_200ps}, one can already see that just with the increased melting temperature of the potential, the ablation process is delayed and the amount of ablated material is reduced. This is due to the bond hardening that the IP reproduces at high degrees of excitation \cite{Kuemmel_2025_MDT}. Also taking the energy conservation into account, the overall electron temperature is lower than in all other simulations, the fluence was insufficient for ablation and the surface just starts melting. Using the additional blast force, it can again be observed that single atoms or small clusters are being removed very shortly after laser irradiation. This happens due to the high pressure gradients as a consequence of the high electron temperature gradients previously shown in figure \ref{fig:press_ANA_vs_ANATE_5ps} and discussed above. In the temporal and spatial evolution of the pressure presented in figure \ref{fig:press_ANA_vs_ANATE_200ps}, more details of the atoms removed from the surface are revealed. In particular, it can be seen that in all simulations, single atoms or small clusters are removed that were only visible in the density plots of the simulation using the additional blast force. \\
In the simulation performed with the additional blast force, the incident fluence was barely not high enough to remove the upper layer of the surface which sticks back to the rest of the sample at the end of the simulation. In all simulations, the ingoing pressure wave is the strongest if the blast force or an electron temperature-dependent potential without energy conservation is used. As previously discussed, the energy conservation scheme leads to a weaker pressure wave. \\
An analysis of the exact amounts of ablated material are beyond this investigation and will be discussed in a further publication where particular focus shall be put on the ballistic electron motion and absorption models.

\section{Summary and conclusion}

In this work, an implementation of electron temperature-dependent potentials was presented. Approaches to how to enforce energy conservation and correctly treat pressure gradients in electron temperature gradients have been presented, discussed and compared to the more commonly-used blast force as an additional force contribution. \\
Large-scale TTM-MD simulations of a copper sample irradiated by a short laser pulse have been presented in which the temperature, pressure and density evolution have been investigated. Using these simulations, it could be confirmed that electron temperature-dependent potentials for copper can delay the melting and ablation process which is a consequence of the bond hardening that occurs in copper at high degrees of excitation. Furthermore, enforcing energy conservation further decreases the electron temperatures that can be reached during the simulation which further delays melting and ablation. \\
In a further publication, the exact amounts of ablated material will be investigated using the here-presented models while also taking ballistic electron motion into account.

\section*{Data availability}
All data that was produced for this study are openly available at \url{https://doi.org/10.18419/DARUS-5743}.

\section*{Code availability}
The source code that was used for the simulations performed in this work is openly available at \url{https://github.com/simon-kuemmel/IMD}.

\section*{Acknowledgment}
This work was funded by the Deutsche Forschungsgemeinschaft (DFG, German Research Foundation) – 461606136. The authors would like to thank the state of Baden-Württemberg for the support of the bwHPC and the DFG for funding under the reference “INST 39/1232-1 FUGG” (bwForCluster NEMO 2).

\section*{CRediT authorship contribution statement}
Simon Kümmel: Conceptualization, Data Curation, Formal Analysis, Investigation, Methodology, Resources, Software, Validation, Visualization, Writing – original draft \\
Johannes Roth: Conceptualization, Funding Acquisition, Project Administration, Resources, Supervision, Validation, Writing – review \& editing

\appendix
\section{Edge cases in energy conservation scheme}
\label{sec:Edge_cases}

If an FD cell initially has an electron temperature above k$_{\text{B}}T_{\text{e}} = \SI{1.2}{eV}$ and looses some of its energy so that the final electron temperature is below \SI{1.2}{eV}, the energy conservation reads
\begin{align}
  & E_{\text{e}} \left( \text{k}_{\text{B}} T_{\text{e}}^{\text{f}} \right) + E_{\text{i}} \left( \text{k}_{\text{B}} T_{\text{i}}^{\text{f}} \right) - \left[ E_{\text{e}} \left( \text{k}_{\text{B}} T_{\text{e}}^{1} \right) + E_{\text{i}} \left( \min \left( \SI{1.2}{eV}, \text{k}_{\text{B}} T_{\text{i}}^{0} \right) \right) \right] \nonumber \\
  \overset{\hphantom{T_{\text{e}}^{\text{f}} = T_{\text{i}}^{\text{f}} \eqcolon T^{\text{f}}}}{=}& \frac{\gamma}{2} \left( \text{k}_{\text{B}} T_{\text{e}}^{\text{f}} \right)^{2} + \lambda \text{k}_{\text{B}} T_{\text{e}}^{\text{f}} + a^{\text{str}} + b \text{k}_{\text{B}} T_{\text{i}}^{\text{f}} + c \left( \text{k}_{\text{B}} T_{\text{i}}^{\text{f}} \right)^{2} \nonumber \\
  &- \frac{\gamma}{2} \left( \text{k}_{\text{B}} T_{\text{e}}^{1} \right)^{2} - \lambda \text{k}_{\text{B}} T_{\text{e}}^{1} - a^{\text{str}} \nonumber \\
  &- b \cdot \min \left( \SI{1.2}{eV}, \text{k}_{\text{B}} T_{\text{i}}^{0} \right) - c \cdot \min \left( \SI{1.2}{eV}, \text{k}_{\text{B}} T_{\text{i}}^{0} \right)^{2} \nonumber \\
  \overset{T_{\text{e}}^{\text{f}} \eqcolon T^{\text{f}}, T_{\text{i}}^{\text{f}} = \SI{1.2}{eV}}{=}& \left( \text{k}_{\text{B}} T^{\text{f}} \right)^{2} \underbrace{\frac{\gamma}{2}}_{\eqcolon A} + \text{k}_{\text{B}} T^{\text{f}} \underbrace{\lambda}_{\eqcolon B} \nonumber \\
  &+ \left[ b \cdot \SI{1.2}{eV} + c \cdot \left( \SI{1.2}{eV} \right)^2 - \frac{\gamma}{2} \left( \text{k}_{\text{B}} T_{\text{e}}^{1} \right)^{2} - \lambda \text{k}_{\text{B}} T_{\text{e}}^{1} \right. \nonumber \\
  &\underbrace{+ \left. b \cdot \min \left( \SI{1.2}{eV}, \text{k}_{\text{B}} T_{\text{i}}^{0} \right) - c \cdot \min \left( \SI{1.2}{eV}, \text{k}_{\text{B}} T_{\text{i}}^{0} \right)^{2} \right]}_{\eqcolon \tilde{C}} . \nonumber \\
  \overset{\hphantom{T_{\text{e}}^{\text{f}} = T_{\text{i}}^{\text{f}} \eqcolon T^{\text{f}}}}{\overset{!}{=}}& \Delta E
  \label{eq:Energy_balance_2}
\end{align}

\section{Local order parameter}
\label{sec:LOP}

\begin{figure}[H]
  \centering
  \begin{subfigure}{0.5 \textwidth}
    \includegraphics[width=0.95 \textwidth]{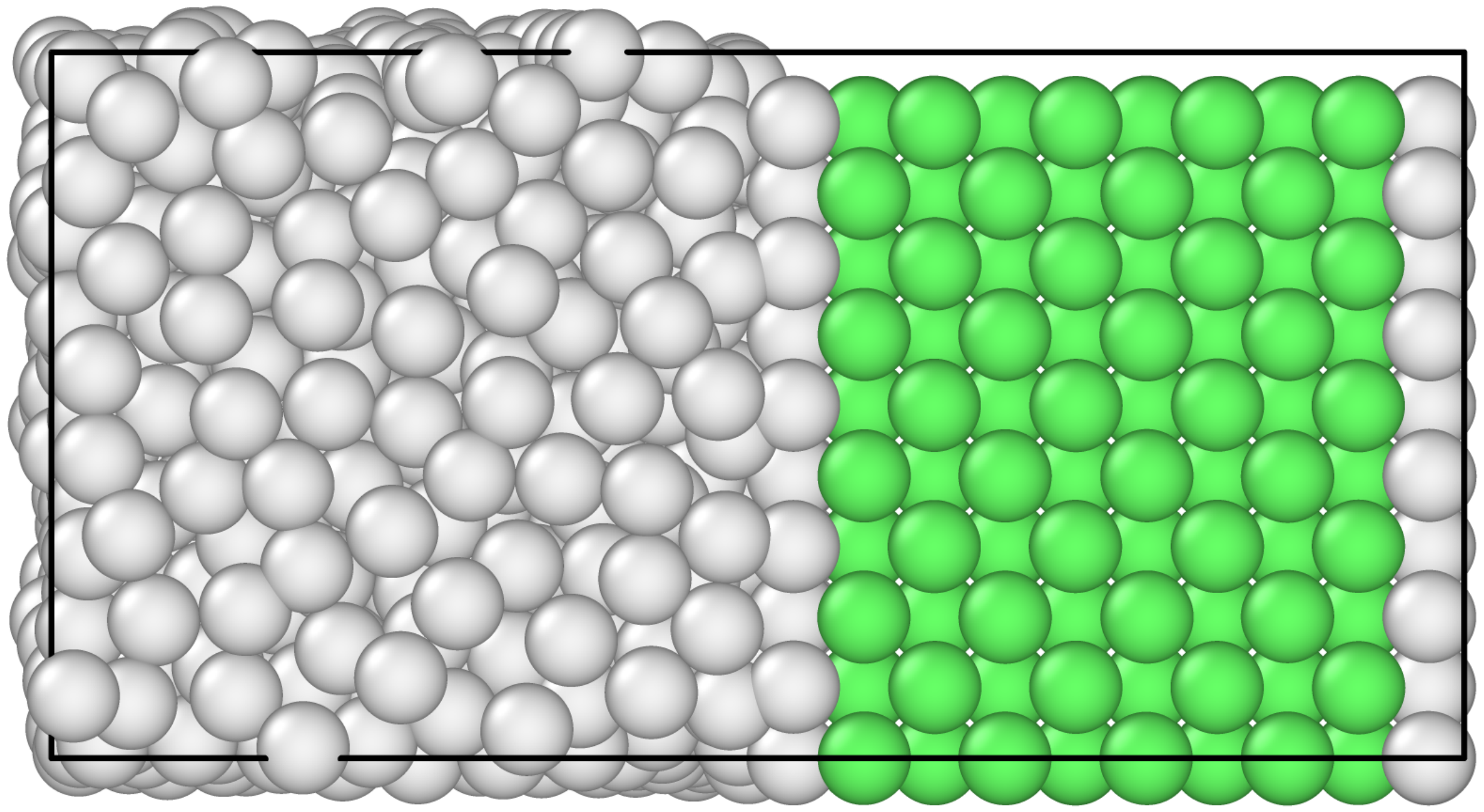}
  \end{subfigure}
  \hfill
  \begin{subfigure}{0.4 \textwidth}
    \includegraphics[width=0.95 \textwidth]{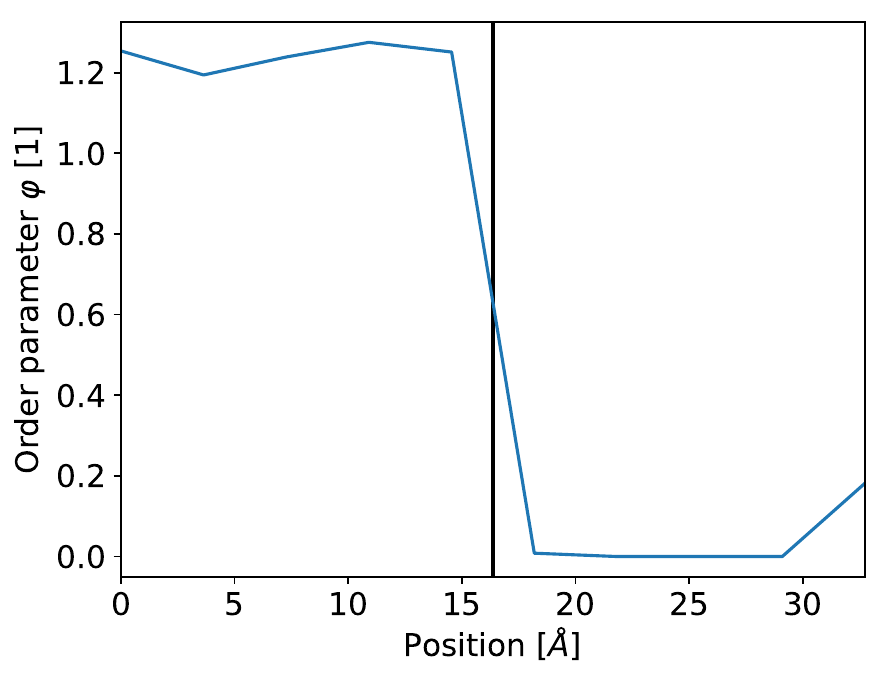}
  \end{subfigure}
  \caption{Left: Simulation setup for the comparison of the LOP calculation using the method described in \cite{Hoyt_2001_LOP} and as implemented for this work. Green atoms are in an FCC structure, white atoms in a disordered structure. This was analyzed using the adaptive common neighbour analysis \cite{Stukowski_2012_CNA} implemented in ovito \cite{Stukowski_2010_ovito}. Right: LOP of the initial configuration for the LOP comparison.}
  \label{fig:LOP_setup}
\end{figure}

\begin{figure}[H]
  \centering
  \begin{subfigure}{0.49 \textwidth}
    \includegraphics[width=0.95 \textwidth]{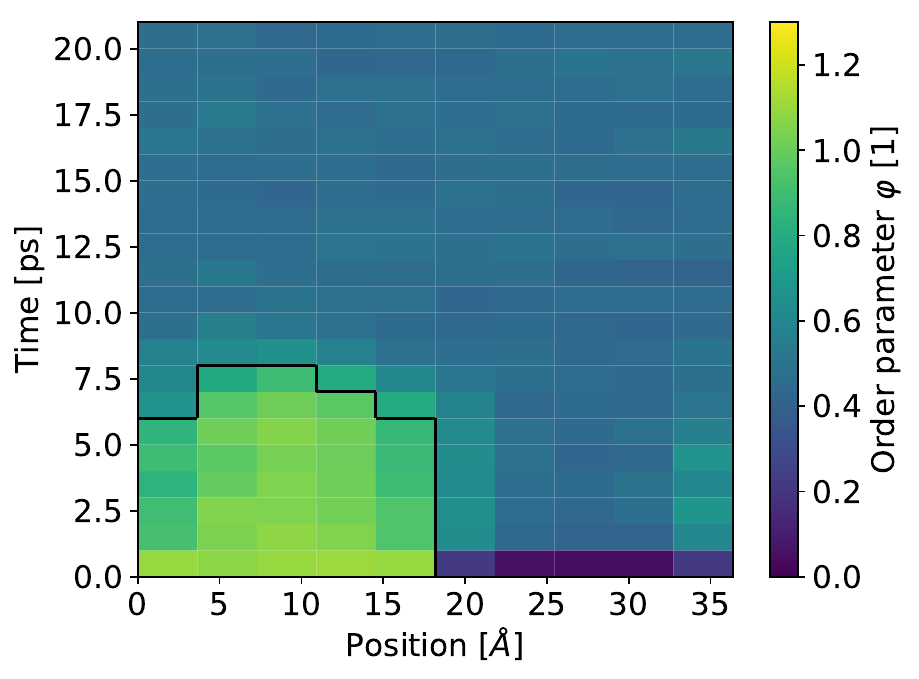}
  \end{subfigure}
  \hfill
  \begin{subfigure}{0.49 \textwidth}
    \includegraphics[width=0.95 \textwidth]{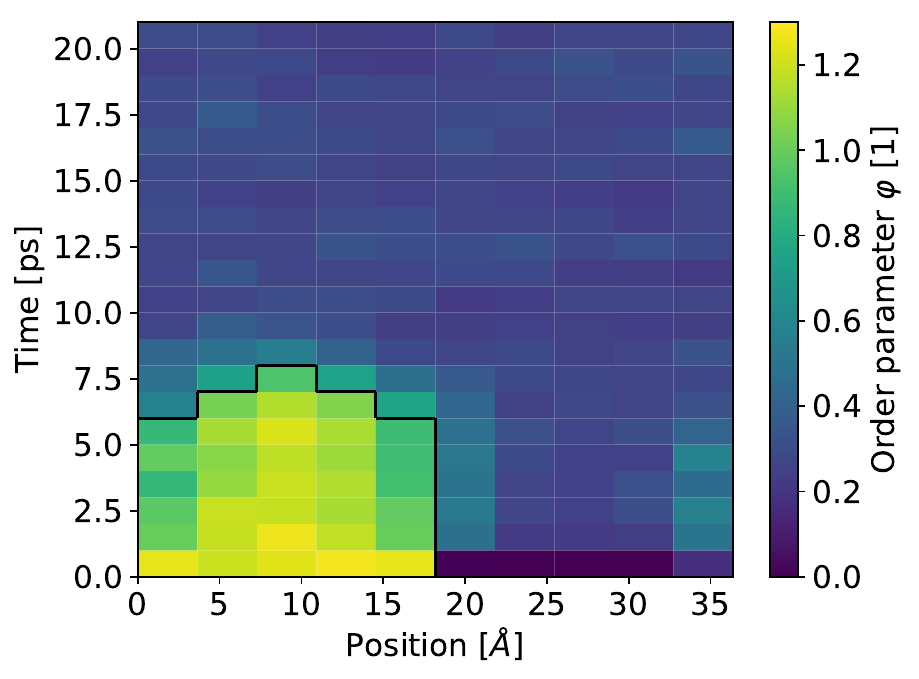}
  \end{subfigure}
  \caption{Left: LOP evolution calculated as described in \cite{Hoyt_2001_LOP}. Right: The same but as implemented for the presented simulations. In both plots, the black lines indicate the solid-liquid transition based on a threshold order parameter. This threshold value is \num{0.7}.}
  \label{fig:order_parameter_diagram}
\end{figure}

In order to investigate the LOP of a partially melted sample, a simulation setup as shown in the left part of figure \ref{fig:LOP_setup} is used. A simulation box with a total of 2000 atoms is created where one half is initially in an ideal FCC structure and the other half is a completely melted structure with the same density. PBCs are applied in all directions. As can be seen in the initial LOP at various locations of the initial configuration shown in the right part of figure \ref{fig:LOP_setup}, the two laves of the complete structure can easily be distinguished. In the solid part, the LOP is close to zero and in the melted part, it is above \num{1.0}. At either end of the sample, the LOP deviates a bit due to PBCs. \\
At a constant temperature of \SI{1625}{K}, the LOP is monitored for \num{20000} timesteps of \SI{1}{fs} in an NVT simulation. The resulting evolution of the order parameter is shown in figure \ref{fig:order_parameter_diagram}. In the left part of that figure, the LOP was calculated using the 12 closest neighbours in a post-processing step as described in \cite{Hoyt_2001_LOP}. In the right part, it was calculated as implemented in for this work using the neighbours within the cutoff distance from equation \ref{eq:d_cutoff}. \\
The resulting LOP evolutions are shown in figure \ref{fig:order_parameter_diagram}. As can be seen there, there's little qualitative difference when only considering the neighbouring atoms within the radius $d_{\text{cutoff}}$ from equation \ref{eq:d_cutoff}. This adapted approach to getting the LOP is therefore considered to be usable.

\bibliographystyle{IEEEtran}
\addcontentsline{toc}{chapter}{Bibliography}
\bibliography{config/references.bib}

\end{document}

%% file: config/config.tex
\usepackage{a4wide}
\usepackage{fancyhdr}  
\usepackage[english]{babel}
\usepackage{upgreek}		

\usepackage[hidelinks]{hyperref}
\usepackage[acronym, nonumberlist, nopostdot]{glossaries}		
\usepackage{float}			
\usepackage[toc,page]{appendix}

\usepackage{tikz}		
\usepackage{pgfplots}
\pgfplotsset{compat=1.5}	

\usepackage{pdfpages}		

\usepackage{caption}		
\usepackage{subcaption}

\usepackage{mathtools}
\usepackage{physics}		
\usepackage{siunitx}
\usepackage{braket}			

\usepackage{multirow}		

\usepackage[affil-it]{authblk}

%% file: config/acronyms.tex
\newacronym{dft}{DFT}{Density Functional Theory}
\newacronym{dos}{DOS}{Density of States}

\newacronym{dcp}{DCP}{Drude Critical Point}

\newacronym{ttm}{TTM}{Two-Temperature Model}
\newacronym{tmm}{TMM}{Transfer Matrix Method}
\newacronym{fd}{FD}{Finite Differences}

\newacronym{md}{MD}{Molecular Dynamics}
\newacronym{pbc}{PBC}{Periodic Boundary Condition}
\newacronym{eam}{EAM}{Embedded Atom Method}
\newacronym{ip}{IP}{Interaction Potential}

\newacronym{fcc}{FCC}{Face-Centered Cubic}

\newacronym{cna}{CNA}{Common Neighbour Analysis}
\newacronym{csp}{CSP}{Centrosymmetry Parameter}
\newacronym{lop}{LOP}{Local Order Parameter}

\newacronym{eos}{EOS}{Equation of State}
\newacronym{flt}{FLT}{Fermi liquid theory}
\newacronym{tfm}{TFM}{Thomas-Fermi model}

\newacronym{fwhm}{FWHM}{Full Width at Half Maximum}

%% file: config/commands.tex
\newcommand{\at}{\left. \vphantom{\frac{1}{1}} \right|}	
\newcommand{\iu}{{i\mkern1mu}}							

\renewcommand{\tableautorefname}{table}
\renewcommand{\figureautorefname}{figure}
\renewcommand{\appendixautorefname}{appendix}
\renewcommand{\sectionautorefname}{section}

\let\originalleft\left		
\let\originalright\right
\renewcommand{\left}{\mathopen{}\mathclose\bgroup\originalleft}
\renewcommand{\right}{\aftergroup\egroup\originalright}